\documentclass{article}

\usepackage[affil-it]{authblk}
\usepackage[usenames,dvipsnames]{xcolor}
\usepackage{amsfonts}
\usepackage{amsmath,amsthm,amssymb,dsfont}
\usepackage{enumerate}
\usepackage[english]{babel}
\usepackage{graphicx}	
\usepackage[caption=false]{subfig}
\usepackage[margin=2.8cm]{geometry}
\usepackage{url}
\usepackage{todonotes}
\usepackage{bbm}
\usepackage{centernot}
\usepackage{blkarray}

\usepackage{tikz,pgfplotstable}
\usetikzlibrary{chains}
\usetikzlibrary{fit}
\usetikzlibrary{spy}
\usepackage{pgflibraryarrows}		
\usepackage{pgflibrarysnakes}		

\usepackage{makecell}

\usepackage{epsfig}
\usetikzlibrary{shapes.symbols,patterns} 
\usepackage{pgfplots}

\usepackage{mathrsfs}

\usepackage{hyperref}
\hypersetup{colorlinks=true,citecolor=blue,linkcolor=blue,filecolor=blue,urlcolor=blue,breaklinks=true}

\usepackage{nicefrac}
\usepackage{mathtools}
\usepackage{xparse}
\usepackage{subfig}

\theoremstyle{plain}
\newtheorem{theorem}{Theorem}[section]
\newtheorem{lemma}[theorem]{Lemma}

\newtheorem{proposition}[theorem]{Proposition}

\theoremstyle{definition}
\newtheorem{definition}[theorem]{Definition}
\newtheorem{remark}[theorem]{Remark}
\newtheorem{example}[theorem]{Example}

\newcommand*{\cA}{\mathcal{A}}
\newcommand*{\cB}{\mathcal{B}}
\newcommand*{\cC}{\mathcal{C}}

\newcommand*{\cE}{\mathcal{E}}

\newcommand*{\cH}{\mathcal{H}}
\newcommand*{\cL}{\mathcal{L}}
\newcommand*{\cI}{\mathcal{I}}
\newcommand*{\cK}{\mathcal{K}}

\newcommand*{\cP}{\mathcal{P}}

\newcommand*{\cW}{\mathcal{W}}
\newcommand*{\cX}{\mathcal{X}}
\newcommand*{\cY}{\mathcal{Y}}

\newcommand*{\RR}{\mathbb{R}}

\newcommand*{\CC}{\mathbb{C}}

\newcommand*{\NN}{\mathbb{N}}

\newcommand*{\id}{I}

\newcommand*{\supp}{\mathrm{supp}}
\newcommand*{\ket}[1]{| #1 \rangle}
\newcommand*{\bra}[1]{\langle #1 |}

\newcommand*{\Pos}{\mathscr{P}}
\newcommand*{\Lin}{\mathscr{L}}
\newcommand*{\Her}{\mathscr{H}}

\newcommand*{\D}{\mathscr{D}}

\newcommand{\imD}[1]{D_{(#1)}}  	
\newcommand{\imH}[1]{H_{(#1)}} 
\newcommand{\imQ}[1]{Q_{(#1)}}
\newcommand{\imHup}[1]{\imH{#1}^\uparrow}
\newcommand{\imQup}[1]{\imQ{#1}^\uparrow}
\newcommand{\imQdi}[1]{\imQ{#1}^{\mathrm{DI}}}

\newcommand{\petD}[1]{\overline{D}_{#1}} 			

\newcommand{\petQ}[1]{\overline{Q}_{#1}}

\newcommand{\swD}[1]{\widetilde{D}_{#1}}  	
\newcommand{\swH}[1]{\widetilde{H}_{#1}} 

\newcommand{\swHup}[1]{\swH{#1}^\uparrow}

\newcommand{\geoD}[1]{\widehat{D}_{#1}}  	

\newcommand{\mesD}[1]{D^{\mathbb{M}}_{#1}}

\newcommand{\strat}{\Sigma}

\newcommand{\tr}[1]{\mathrm{Tr}\left[#1\right]} 
\newcommand{\ptr}[2]{\mathrm{Tr}_{#1}\left[#2\right]}

\let\inner\relax
\NewDocumentCommand\inner{mg}{%
	\ensuremath{\left\langle #1, \! \IfNoValueTF{#2}{#1}{#2}\right\rangle}%
}
\let\outer\relax
\NewDocumentCommand\outer{mg}{%
	\ensuremath{\ket{#1}\!\! \IfNoValueTF{#2}{\bra{#1}}{\bra{#2}}}%
}

\begin{document}

\title{{\LARGE Computing conditional entropies for quantum correlations}}

\author[1]{Peter Brown}
\author[2]{Hamza Fawzi}
\author[1]{Omar Fawzi}

\affil[1]{\small{Univ Lyon, ENS Lyon, UCBL, CNRS,  LIP, F-69342, Lyon Cedex 07, France}}
\affil[2]{\small{DAMTP, University of Cambridge, United Kingdom}}

\maketitle

\begin{abstract}
The rates of quantum cryptographic protocols are usually expressed in terms of a conditional entropy minimized over a certain set of quantum states. In particular, in the device-independent setting, the minimization is over all the quantum states jointly held by the adversary and the parties that are consistent with the statistics that are seen by the parties. Here, we introduce a method to approximate such entropic quantities. Applied to the setting of device-independent randomness generation and quantum key distribution, we obtain improvements on protocol rates in various settings. In particular, we find new upper bounds on the minimal global detection efficiency required to perform device-independent quantum key distribution without additional preprocessing. Furthermore, we show that our construction can be readily combined with the entropy accumulation theorem in order to establish full finite-key security proofs for these protocols. 

In order to achieve this we introduce the family of \emph{iterated mean} quantum R\'enyi divergences with parameters $\alpha_k = 1+\frac{1}{2^{k}-1}$ for positive integers $k$. We then show that the corresponding conditional entropies admit a particularly nice form which, in the context of device-independent optimization,  can be relaxed to a semidefinite programming problem using the Navascu\'es-Pironio-Ac\'in hierarchy. 
\end{abstract}

\section{Introduction}


Quantum cryptography is one of the most promising applications in the field of emerging quantum technologies having already seen commercial implementations. Using quantum systems it is possible to execute cryptographic protocols with security based on physical laws~\cite{BB84} -- as opposed to assumptions of computational hardness. To date, much progress has been made in the development of new protocols and their respective security proofs. However, in real world implementations such protocols are not infallible. Side-channel attacks arising from hardware imperfections or unreasonable assumptions in the security analysis can render the protocols useless~\cite{LWWESM}. Whilst improvements in the hardware and more detailed security analyses can fix these issues, quantum theory also offers an alternative approach: device-independent (DI) cryptography.

Pioneered by the work of~\cite{MayersYao}, device-independent cryptography circumvents the majority of side-channel attacks by offering security whilst making minimal assumptions about the hardware used in the protocol. Typically, the devices used within an implementation of a DI protocol are treated as black boxes. The remarkable fact that one can still securely perform certain cryptographic tasks on untrusted devices is a consequence of Bell-nonlocality~\cite{Bell}. In short, if an agent observes nonlocal correlations between two or more devices then they can infer restrictions on the systems used to produce them. It is then possible for the agent to infer additional desirable properties of their devices by analyzing this restricted class of systems. For example, it is known that all nonlocal correlations are necessarily random~\cite{MAG}. As a consequence, we can construct randomness generation~\cite{ColbeckThesis,CK2,PAMBMMOHLMM} and quantum key distribution (QKD) protocols~\cite{MayersYao,Ekert} with device-independent security. 

A central problem in the development of new DI protocols is the question of how to calculate the \emph{rate} of a protocol. I.e., in DI-RNG how much randomness is generated or in DI-QKD how much secret key is generated per use of the device. 
For many DI protocols, including DI-RNG and DI-QKD, this problem reduces to minimizing the conditional von Neumann entropy over a set of quantum states that are characterized by restrictions on the correlations they can produce. Unfortunately, directly computing such an optimization is a highly non-trivial task. Firstly, conditional entropies are non-linear functions of the states of a system and so the resulting optimization is in general non-convex and a naive optimization is not guaranteed to return a global optima. Moreover, as we are working device-independently we cannot assume any a priori bound on the dimensions of the systems used within the protocol. Nevertheless, in certain special cases the problem can be solved analytically~\cite{PABGMS}. However, the techniques used in the analysis of~\cite{PABGMS} rely on particular algebraic properties of devices with binary inputs and binary outputs. As such, they do not generalize to more complex protocols with more inputs or outputs. This prompts the development of general numerical techniques to tackle this problem. 

Simple numerical lower bounds on the von Neumann entropy minimization can be obtained through the min-entropy~\cite{KRS}. It was shown in~\cite{BSS14,NPS14} that the analogous optimization of the min-entropy can be expressed as a noncommutative polynomial of measurement operators. This problem can then be relaxed to a semidefinite program (SDP) using the NPA hierarchy~\cite{NPA_general} which can then be solved efficiently. This approach gives a simple and efficient method to lower bound the rates of various DI tasks and has found widespread use in the analysis of DI protocols. Unfortunately, the min-entropy is in general much smaller than the von Neumann entropy and so this approach usually produces suboptimal results. More recently, the authors of~\cite{TSGPL19} extended the work of~\cite{CML16} to the device-independent setting. By viewing the objective function as an entropy gain between the systems producing the correlations they were able to construct a method to derive a noncommutative polynomial of the measurement operators that lower bounds the conditional von Neumann entropy. As in the case of the min-entropy approach, this can be approximated efficiently by an SDP. The numerical results presented in~\cite{TSGPL19} are very promising, providing significant improvements in the rates when compared to the min-entropy approach and also improving over the analytical results of~\cite{PABGMS}. However, their approach is relatively computationally intensive requiring the optimization of a degree $6$ polynomial in the simplest setting. For comparison, in protocols involving two devices the min-entropy can always be computed using a polynomial of degree no larger than $2$.

\subsection{Contributions of this work}

In this work we take a different approach, defining a new family of quantum R\'enyi divergences, the \emph{iterated mean divergences}.
The iterated mean divergences are defined as solutions to certain SDPs and their constructions are inspired by the semidefinite representations of the weighted matrix geometric means~\cite{FS17}. We call them quantum R\'enyi divergences as they match for commuting operators with the classical R\'enyi divergence. As R\'enyi divergences are well studied objects in information theory and have found numerous operational interpretations the iterated mean divergences may also be of independent interest. In fact, our new divergences have already inspired the definition of other quantum divergences with different information-theoretic applications~\cite{FF20}. The key property of iterated mean divergences that makes them suited for DI optimization is that their SDP representation does not explicitly refer to the dimension of the underlying quantum systems. With this property, the corresponding conditional entropy of a state $\rho$ can be written as a maximization of a noncommutative Hermitian polynomial in some operators $V_1, \dots, V_m$ evaluated on the state $\rho$ and the operators $V_1, \dots, V_m$ are subject to polynomial inequalities that are dimension independent. We refer to Remark~\ref{rem:dimension-free} for a more detailed discussion of this point.

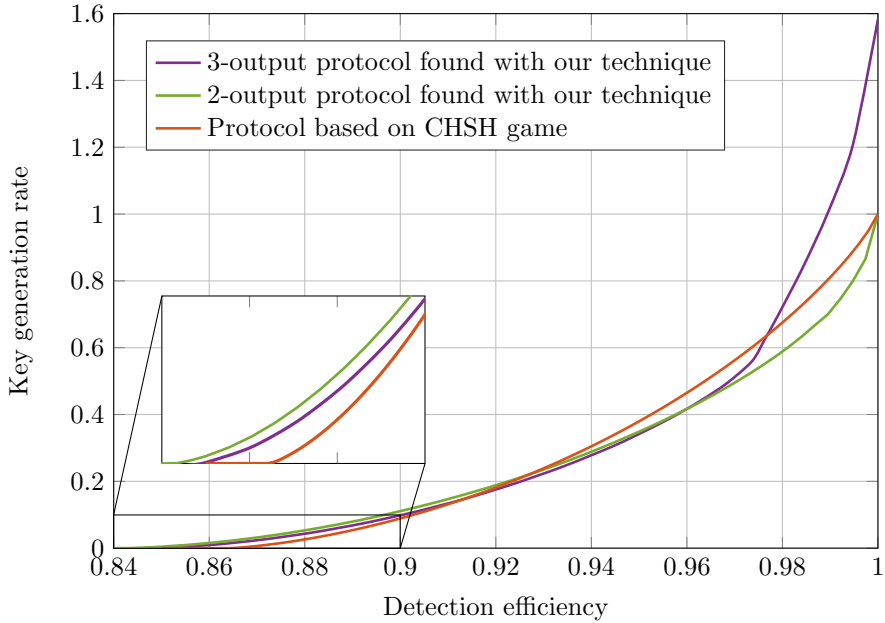
\begin{figure}[t!]
	\centering
	\definecolor{mycolor2}{rgb}{0.00000,0.44700,0.74100}%
	\definecolor{mycolor4}{rgb}{0.85000,0.32500,0.09800}%
	\definecolor{mycolor3}{rgb}{0.92900,0.69400,0.12500}%
	\definecolor{mycolor1}{rgb}{0.49400,0.18400,0.55600}%
	\definecolor{mycolor5}{rgb}{0.4660, 0.6740, 0.1880}%
	\begin{tikzpicture}[scale=1.0]
	
	\begin{axis}[%
	width=4in,
	height=2.8in,
	scale only axis,
	xmin=0.84,
	xmax=1.0,
	ymin=0,
	ymax=1.6,
	grid=major,
	xlabel={Detection efficiency},
	ylabel={Key generation rate},
	xtick={0.84, 0.86, 0.88, 0.9, 0.92, 0.94, 0.96, 0.98, 1},
	axis background/.style={fill=white},
	legend style={at={(0.8,0.95)},legend cell align=left, align=left, draw=white!15!black}
	]
	\addplot[smooth, color=mycolor1, line width=1.pt] table[col sep=comma] {./h43_optbin_qutrits_tikz.dat};
	\addlegendentry{3-output protocol found with our technique}
	\addplot[ color=mycolor5, line width=1.pt] table[col sep=comma] {./h87_de_optbin_tikz.dat};
	\addlegendentry{2-output protocol found with our technique}
	\addplot[ color=mycolor4, line width=1.pt] table[col sep=comma] {./h_rates_tikz.dat};
	\addlegendentry{Protocol based on CHSH game}
	\coordinate (insetPosition) at (axis cs:0.85,0.2);
	\coordinate (pos1big) at (axis cs:0.84,0.0);
	\coordinate (pos2big) at (axis cs:0.9,0.1);
	\coordinate (topleftbig) at (axis cs:0.84,0.1);
	\coordinate (bottomrightbig) at(axis cs:0.9,0.0);
	\end{axis}
	\begin{axis}[at={(insetPosition)},anchor={outer south west},
	width=2in,
	height=1.5in,
	xmin=0.84,
	xmax=0.9,
	ymin=0,
	ymax=0.1,
	xtick = {0.86, 0.88},
	xticklabels={,,},
	ymajorticks=false,
	axis background/.style={fill=white}
	]
	\addplot[smooth, color=mycolor1, line width=1.2pt] table[col sep=comma] {./h43_optbin_qutrits_tikz.dat};
	\addplot[ color=mycolor5, line width=1.pt] table[col sep=comma] {./h87_de_optbin_tikz.dat};
	\addplot[ smooth, color=mycolor4, line width=1.2pt] table[col sep=comma] {./h_rates_tikz.dat};
	\coordinate (topleftsmall) at (axis cs:0.84,0.1);
	\coordinate (bottomrightsmall) at (axis cs:0.9,0.0);
	\end{axis}
	\draw (pos1big) rectangle (pos2big); 
	\draw (topleftbig) -- (topleftsmall);
	\draw (bottomrightbig) -- (bottomrightsmall);
	\end{tikzpicture}%
	\caption{Comparison of asymptotic key rates (without preprocessing) for a DI-QKD protocol based on the CHSH game, a DI-QKD protocol for 2-input 2-output devices and a DI-QKD protocol for 2-input 3-output devices. The key rates for the 3-output protocol can be smaller than the 2-output protocol in the regime of high noise as they were evaluated using different entropies from our family. However, for low detection efficiency we see that the 3-output protocol can achieve key rates of up to $\log(3)$ bits.}
	\label{fig:qkd_comparison_simple}
\end{figure}

We then apply our divergences to the task of computing rates of DI randomness expansion (RE) and DI-QKD protocols. We compare the rates certified by our techniques with those certified by the min-entropy, the method of~\cite{TSGPL19} and an analytical bound on $H(A|E)$ derived for the CHSH game~\cite{PABGMS}. Compared to the min-entropy bound, as will be shown in the examples we consider throughout the paper, our method almost always gives a significantly improved bound at a minor additional computational cost. Compared to the known analytical bound for CHSH, our method can be applied to a large family of protocols and this allows us to \emph{search} for protocols that improve the various properties of interest. For example, by optimizing over a family of protocols with two inputs and two outputs per device we find a new upper bound on the minimal detection efficiency required to perform DI-QKD with a two-qubit system without additional preprocessing (see Fig.~\ref{fig:qkd_comparison_simple}).\footnote{Very recently, it was shown in~\cite{HSTRBS20,WAP20} that noisy preprocessing of the raw key could reduce the minimal detection efficiency required for a protocol based on the CHSH game~\cite{PABGMS}. Here we do not take this preprocessing into account but it would be interesting to explore whether the same ideas could be combined with our numerical methods to further reduce the detection efficiency thresholds.}

Compared to the numerical method of~\cite{TSGPL19}, a major advantage of our new method is in its simplicity and flexibility in several respects. First, the noncommutative polynomial optimization problems that we construct are of degree at most 3 regardless of the number of inputs and outputs of the devices whereas for~\cite{TSGPL19} the degree is six for the smallest possible setting and it grows with the number of inputs and outputs. Second, the coefficients appearing in the SDPs are explicit small integers for our method whereas for~\cite{TSGPL19}, they involve closed form solutions to integrals of the $\beta$ functions appearing in the multivariate trace inequality of~\cite{sutter2017multivariate}. Third, our method is flexible to use as it has a parameter $k \in \NN$ that can be increased to improve the bounds at the cost of increasing the size of the resulting SDP. This simplicity and flexibility is what allowed us to find the improved protocol described in Fig.~\ref{fig:qkd_comparison_simple} for example. We also illustrate the versatility of our method by applying it to more exotic settings like randomness certification using sequential measurements~\cite{BBS}. In terms of the numerical bounds obtained, we found improvements compared to~\cite{TSGPL19} in the low noise regime but for high noise, the method of~\cite{TSGPL19} produced higher rates. However, we only perform such a direct comparison for the smallest setting of 2-input 2-output devices as no data for larger setups is provided for the technique of~\cite{TSGPL19} in its full generality.

Finally, we demonstrate that our method can be used directly with the entropy accumulation theorem~\cite{DFR, DF} by constructing explicit min-tradeoff functions from the solutions of our optimizations. Applying the security proof blueprints developed in~\cite{ARV,ADFRV} our techniques can readily be used together with the entropy accumulation theorem to construct complete security proofs of many DI protocols. This property is again due to the simplicity of our method and it is unclear whether this can be done with other numerical methods such as the one in~\cite{TSGPL19}. For all these reasons, we anticipate that the numerical tools developed here will lead to the development of better device-independent protocols.

The rest of the paper is structured as follows. We begin by defining the iterated mean divergences and then prove numerous properties including several dual formulations, a data-processing inequality and we relate to other notions of quantum R\'enyi divergences. We then demonstrate how to apply these divergences to the task of computing the rates of different DI protocols, giving several examples and comparing the results to the techniques of~\cite{TSGPL19} and the CHSH based protocol of~\cite{PABGMS}. Finally, we conclude with several avenues for further research. 


\section{Preliminaries}

We define $\NN$ to be the set of strictly positive integers. Let $\cH$ be a Hilbert space; we denote the set of linear operators on $\cH$ by $\Lin(\cH)$, the set of Hermitian operators on $\cH$ by $\Her(\cH)$, the set of positive semidefinite operators on $\cH$ by $\Pos(\cH)$ and the set of positive semidefinite operators with unit trace on $\cH$ by $\D(\cH)$. All Hilbert spaces in this work are finite dimensional unless otherwise stated. Given a linear map $\cE : \Lin(\cH_1) \rightarrow \Lin(\cH_2)$, we say $\cE$ is CPTP if it is completely positive and trace preserving. Given two Hilbert spaces $\cH$ and $\cK$ we write $\cH\cK$ as shorthand for $\cH\otimes \cK$. Given two operators $A,B \in \Lin(\cH)$ we write $A \leq B$ if $B-A \in \Pos(\cH)$. The support of an operator $A \in \Lin(\cH)$, denoted $\supp(A)$, is the orthogonal complement of its kernel, $\mathrm{ker}(A) = \{x \in \cH : Ax = 0\}$. For $A,B \in \Lin(\cH)$, we write $A \ll B$ if $\supp(A)  \subseteq \supp(B)$. For $A \in \Lin(\cH)$, $A^*$ denotes its adjoint and if $A$ is nonsingular then $A^{-1}$ denotes its inverse. If $A$ is singular then $A^{-1}$ denotes the Moore-Penrose pseudo-inverse of $A$. We use the symbol $\id$ to denote the identity operator. A collection of operators $\{M_1, \dots, M_n\}$ forms an $n$-outcome POVM on $\cH$ if $\sum_{i=1}^n M_i = \id$ and $M_i \in \Pos(\cH)$ for all $i = 1,\dots, n$. 

The geometric mean of two positive definite matrices $A$ and $B$ is defined as $$A \# B = A^{1/2}(A^{-1/2} B A^{-1/2})^{1/2} A^{1/2}.$$ This definition can be extended to positive semidefinite matrices $A, B$ as $\lim_{\epsilon \to 0} A_{\epsilon} \# B_{\epsilon}$ where $X_{\epsilon} = X + \epsilon \id$. The geometric mean has the property that if $C \leq D$ then $A \# C \leq A \# D$~\cite[Corollary 3.2.3]{hiainotes}.

Let $\alpha \in (0,1) \cup (1, \infty)$, $\rho \in \D(\cH)$ and $\sigma \in \Pos(\cH)$ with $\rho \ll \sigma$. The \emph{Petz-R\'enyi divergence}~\cite{Petz} of order $\alpha$ is defined as
\begin{equation}
\petD{\alpha}(\rho \| \sigma) := \frac{1}{\alpha - 1} \log \tr{ \rho^\alpha \sigma^{1-\alpha}}.
\end{equation}
The \emph{sandwiched R\'enyi divergence}~\cite{MDSFT,WWY13} of order $\alpha$ is defined as
\begin{equation}
\swD{\alpha}(\rho \| \sigma) := \frac{1}{\alpha - 1} \log \tr{\left(\sigma^{\frac{1-\alpha}{2\alpha}} \rho \sigma^{\frac{1-\alpha}{2\alpha}}\right)^\alpha}.
\end{equation}
In the limit $\alpha \to 1$ both the Petz-R\'enyi divergence and the sandwiched R\'enyi divergence converge to the \emph{Umegaki relative entropy}~\cite{umegaki}
\begin{equation}
D(\rho \|\sigma) := \tr{\rho(\log \rho - \log \sigma)}.
\end{equation}
The \emph{geometric R\'enyi divergence}~\cite{M15}  of order $\alpha$ is defined as
\begin{equation}
\geoD{\alpha}(\rho \| \sigma) := \frac{1}{\alpha -1} \log \tr{\rho^{1/2}\left(\rho^{-1/2} \sigma \rho^{-1/2}\right)^{1-\alpha} \rho^{1/2}}.
\end{equation}
In the limit $\alpha \to 1$ the geometric R\'enyi divergence converges to the \emph{Belavkin-Staszewski relative entropy} $\tr{\rho \log (\rho^{1/2} \sigma^{-1} \rho^{1/2})}$~\cite{BS82}. The geometric R\'enyi divergence is the largest R\'enyi divergence satisfying data-processing. 
The \emph{max divergence} is defined as 
\begin{equation}
D_{\max}(\rho \| \sigma) := \log \inf \{\lambda > 0 : \rho \leq \lambda \sigma\}. 
\end{equation}
Finally, the \emph{measured R\'enyi divergence} is defined as the largest classical divergence obtained from measuring $\rho$ and $\sigma$. For $\alpha \in (1,\infty)$ this is formally defined as
\begin{equation}
\mesD{\alpha}(\rho\| \sigma) := \frac{1}{\alpha - 1} \log \sup_{\{M_i\}_i} \sum_i \tr{M_i \rho}^{\alpha} \tr{M_i \sigma}^{1-\alpha},
\end{equation}
where the supremum is taken over all POVMs $\{M_i\}$. This divergence also admits the following variational characterization~\cite{BFT17}
\begin{equation}
\label{eq:mes_var_expr}
\mesD{\alpha}(\rho\| \sigma) = \frac{1}{\alpha - 1} \log \sup_{\omega > 0} \alpha \tr{\rho \omega^{1-\tfrac{1}{\alpha}}} +  (1-\alpha) \tr{\sigma \omega}. 
\end{equation} 

Given bipartite state $\rho_{AB} \in \D(AB)$ and a R\'enyi divergence $\mathbb{D}$ we define a corresponding conditional entropy 
\begin{equation}
\mathbb{H}^{\downarrow}(A|B)_{\rho} := - \mathbb{D}(\rho_{AB} \| \id_A \otimes \rho_B)
\end{equation}
and a corresponding optimized conditional entropy
\begin{equation}
\mathbb{H}^{\uparrow}(A|B)_{\rho} := \sup_{\sigma_B \in \D(B)} - \mathbb{D}(\rho_{AB} \| \id_A \otimes \sigma_B). 
\end{equation}
The \emph{min-entropy} is defined as
\begin{equation}
H_{\min}(A|B) = \sup_{\sigma_B \in \D(B)} -D_{\max}(\rho_{AB} \| \id_A \otimes \sigma_B).
\end{equation}

\section{Semidefinite programs for the iterated mean divergence}

The main technical contribution of this work is the introduction of a family of R\'enyi divergences that are amenable to device-independent optimization. Throughout the remainder of this work we define the sequence $\alpha_k := 1 + \frac{1}{2^k - 1}$ for $k \in \NN$. We note that the name ``iterated mean'' comes from the expression that we establish later in~\eqref{eq:dual4_main}.

\begin{definition}[Iterated mean divergences]
	Let $\cH$ be a Hilbert space, $\rho \in \D(\cH)$, $\sigma \in \Pos(\cH)$ with $\rho \ll \sigma$ and let $\alpha_k = 1+ \frac{1}{2^k-1}$ for each $ k \in \NN$. Then for each $k \geq 1$ we define the \emph{iterated mean divergence} of order $\alpha_k$ as 
\begin{align}
\label{eq:def_imD}
\imD{\alpha_k}(\rho \| \sigma) := \frac{1}{\alpha_k-1} \log \imQ{\alpha_k}(\rho \| \sigma) \ ,
\end{align}
with
\begin{equation}
\begin{aligned}
\label{eq:def_imQ}
\imQ{\alpha_k}(\rho \| \sigma) := &\max_{V_1, \dots, V_k, Z} \,\,\alpha_k \tr{\rho \frac{(V_1 + V_1^{*})}{2}} - (\alpha_k - 1) \tr{\sigma Z} \\
& \quad\mathrm{s.t.} \quad\,\, V_1 + V_1^* \geq 0 \\
& \quad \quad \,\, \begin{pmatrix} \id & V_1 \\ V_1^* & \frac{(V_2 + V_2^*)}{2} \end{pmatrix} \geq 0 \quad \begin{pmatrix} \id & V_2 \\ V_2^* & \frac{(V_3 + V_3^*)}{2} \end{pmatrix} \geq 0 \quad \cdots \quad \begin{pmatrix} \id & V_k \\ V_k^{*} & Z \end{pmatrix} \geq 0 ,
\end{aligned}
\end{equation}
where the optimization varies over $V_1, \dots, V_k \in \Lin(\cH)$ and $ Z \in \Pos(\cH)$. We may assume further that $Z \ll \sigma$ and $V_i \ll \sigma$ for each $i \in \{1,2, \dots, k\}$. Note that by Schur complement (Lemma~\ref{lem:schur_complement}), we may equivalently write the constraints as
\begin{align*}
V_1 + V_1^* \geq 0 \quad \frac{V_2 + V_2^*}{2} \geq V_1^* V_1 \quad \cdots \quad \frac{V_k + V_{k}^*}{2} \geq V_{k-1}^* V_{k-1} \quad  Z \geq V_k^* V_k \ .
\end{align*}
\end{definition}
\begin{remark}[Important property for device-independent optimization]
\label{rem:dimension-free}
The crucial property that makes these divergences well-adapted for device-independent optimization is the fact that $\imQ{\alpha_k}(\rho \| \sigma)$ has a \emph{free} variational formula as a supremum of linear functions in $\rho$ and $\sigma$. We say that $Q$ has a free variational formula if there exists $m,n \in \NN$ and noncommutative Hermitian polynomials $p_1, \dots, p_n$ in the variables $(V_1, \dots, V_m)$ such that for any dimension $d \geq 1$, $\rho \in \D(\CC^d)$ and $\sigma \in \Pos(\CC^d)$
\begin{align}
\label{eq:free_semi_alg}
Q( \rho \| \sigma) = \max_{(V_1,V_2, \dots, V_m) \in S(d)} \tr{V_1 \rho} + \tr{V_2\sigma} \ ,
\end{align}
where the family of sets $\{S(d)\}_{d \in \NN}$ are all defined using the same polynomials $p_1, \dots, p_n$, i.e., 
\begin{align}
S(d) = \big\{ (V_1, \dots, V_m) \in (\CC^{d \times d})^{m}: p_j(V_1, \dots, V_m) \geq 0 \quad \forall j \in \{1, \dots, n\} \big\} \ .
\end{align}
We repeat that the important property is that the sets $S(d)$ describing the linear functions have a \emph{uniform description} that is independent of the dimension $d$ (the polynomials $p_j$ are the same for all dimensions $d$). Such families of sets are studied in the area of free semialgebraic geometry (see e.g.,~\cite{Net19,HKM16}).
Note that the measured R\'enyi divergences have such a formulation as expressed in~\eqref{eq:mes_var_expr} (for rational values of $\alpha$), but these divergences can be smaller than the Umegaki divergence and thus cannot be used to give lower bounds on the von Neumann entropy. It remains an important open problem whether the quantities $\widetilde{Q}_{\alpha}$ or $\overline{Q}_{\alpha}$ (defined by $\swD{\alpha} = \frac{1}{\alpha - 1} \log \widetilde{Q}_{\alpha}$ and $\petD{\alpha} = \frac{1}{\alpha - 1} \log \overline{Q}_{\alpha}$) have free variational formulas of the form~\eqref{eq:free_semi_alg}. Here, we have introduced new divergences $\imD{\alpha_k}$ that have this property by construction. Note that a representation as in~\eqref{eq:free_semi_alg} immediately establishes joint convexity of $Q$ (regardless of the freeness of the representation). 
As such finding a free variational formula for $\widetilde{Q}_{\alpha}$ or $\overline{Q}_{\alpha}$ would provide a ``dimension-free'' proof of joint convexity and, as $\swD{\alpha}$ and $\petD{\alpha}$ are known to converge to $D$ as $\alpha \to 1$, such free variational formulas would lead to converging approximations for the von Neumann entropy that we aim to approximate.
With the divergences $\imD{\alpha_k}$, we can only guarantee convergence as $k \to \infty$ to the von Neumann entropy in the commuting case. In the general case, it remains open to determine the limit as $k \to \infty$ of $\imD{\alpha_k}$.
\end{remark}
The following proposition details some alternate formulations and properties of the iterated mean divergences. We defer the proof of this proposition to the appendix. 
\begin{proposition}\label{prop:properties}
Let $\rho \in \D(\cH)$, $\sigma \in \Pos(\cH)$ and $k \in \NN$. Then the following all hold:
\begin{enumerate}
\item (Rescaling) 
\begin{equation}\label{eq:primal_rescaled}
\begin{aligned}
\imQ{\alpha_k}(\rho \| \sigma) = \max_{V_1, \dots, V_k, Z}& \,\, \left(\tr{\rho \frac{(V_1 + V_1^{*})}{2}} \right)^{\alpha_k}  \\
\mathrm{s.t.}& \quad  \tr{\sigma Z} = 1 \\
& \quad V_1 + V_1^* \geq 0 \\
& \quad \begin{pmatrix} \id & V_1 \\ V_1^* & \frac{(V_2 + V_2^*)}{2} \end{pmatrix} \geq 0 \quad \begin{pmatrix} \id & V_2 \\ V_2^* & \frac{(V_3 + V_3^*)}{2} \end{pmatrix} \geq 0 \quad \cdots \quad \begin{pmatrix} \id & V_k \\ V_k^{*} & Z \end{pmatrix} \geq 0 \ .
\end{aligned}
\end{equation}
\item (Dual formulations)
We have
\begin{equation}\label{eq:dual2_main}
\begin{aligned}
\imQ{\alpha_k}(\rho \| \sigma) = &\min_{A_1, \dots, A_k, C_1, \dots, C_{k} } \frac{1}{2^k-1} \sum_{i=1}^k 2^{k-i} \tr{A_i} \\
& \qquad \qquad\mathrm{s.t.} \quad C_1 \geq \rho \\
& \qquad \qquad \quad \,\, \begin{pmatrix} A_1 & C_1 \\  C_1 & C_2 \end{pmatrix} \geq 0 \quad \begin{pmatrix} A_2 & C_2 \\ C_2 & C_3 \end{pmatrix} \geq 0 \qquad \cdots \qquad \begin{pmatrix} A_k & C_{k} \\ C_{k} & \sigma \end{pmatrix} \geq 0 \ .
\end{aligned}
\end{equation}
Or also
\begin{equation}\label{eq:dual3_main}
\begin{aligned}
\imQ{\alpha_k}(\rho \| \sigma) &= \min_{A_1, \dots, A_k, C_1, \dots, C_{k} } \tr{A_1} \\
& \qquad \qquad\mathrm{s.t.} \quad \tr{A_1} = \tr{A_2} = \dots = \tr{A_k} \\
& \qquad \qquad \quad \,\, C_1 \geq \rho \\
& \qquad \qquad \quad \,\, \begin{pmatrix} A_1 & C_1 \\  C_1 & C_2 \end{pmatrix} \geq 0 \quad \begin{pmatrix} A_2 & C_2 \\ C_2 & C_3 \end{pmatrix} \geq 0 \qquad \cdots \qquad \begin{pmatrix} A_k & C_{k} \\ C_{k} & \sigma \end{pmatrix} \geq 0 \ .
\end{aligned}
\end{equation}
Finally and eponymously
\begin{equation}\label{eq:dual4_main}
\begin{aligned}
\imQ{\alpha_k}(\rho \| \sigma) &= \min_{A_1, \dots, A_k} \tr{A_1} \\
& \qquad \qquad\mathrm{s.t.} \quad  \tr{A_1} = \tr{A_2} = \dots = \tr{A_k} \\
&\qquad \qquad \qquad \rho \leq  A_1 \# ( A_2 \# (\dots \# (A_k \# \sigma) \dots) ).
\end{aligned}
\end{equation}
\item (Submultiplicativity) Let $\rho_1 \in \D(\cH_1)$, $\sigma_1 \in \Pos(\cH_1)$, $\rho_2 \in \D(\cH_2)$ and $\sigma_2 \in \Pos(\cH_2)$. Then, 
\begin{equation}
\imD{\alpha_k}(\rho_1 \otimes \rho_2 \| \sigma_1 \otimes \sigma_2) \leq \imD{\alpha_k}(\rho_1 \| \sigma_1) + \imD{\alpha_k}( \rho_2 \| \sigma_2)\,.
\end{equation}
\item (Relation to other R\'enyi divergences)
\begin{align}
\label{eq:relation_im_other_div}
D^{\mathbb{M}}_{\alpha_k}(\rho \| \sigma) \leq \widetilde{D}_{\alpha_k}(\rho \| \sigma) \leq \imD{\alpha_k}(\rho \| \sigma) \leq \widehat{D}_{\alpha_k}(\rho \| \sigma)
\end{align}
\item (Decreasing in $k$)
For all $k \geq 2$,
\begin{equation}\label{eq:property-ordering}
\imD{\alpha_k}(\rho \| \sigma) \leq \imD{\alpha_{k-1}}(\rho \| \sigma).
\end{equation}
\item (Data processing)
Let $\cK$ be another Hilbert space and let $\cE: \Lin(\cH) \rightarrow \Lin(\cK)$ be a CPTP map, then 
\begin{equation}
\imD{\alpha_k}(\rho \| \sigma) \geq \imD{\alpha_k}(\cE(\rho) \| \cE(\sigma)).
\end{equation}
\item (Reduction to classical divergence) If $[\rho, \sigma] = 0$ then 
\begin{equation}
\imD{\alpha_{k}}(\rho\|\sigma) = \frac{1}{\alpha_k - 1} \log \tr{\rho^{\alpha_{k}} \sigma^{1- \alpha_{k}}}.
\end{equation}
\end{enumerate}
\end{proposition}

\begin{remark}[Relation to $\petD{2}(\rho\|\sigma)$]
	We can show that $\imD{2}(\rho \| \sigma)$ is no larger than the Petz-R\'enyi divergence $\petD{2}(\rho \| \sigma)$. 
	By the Schur complement (Lemma~\ref{lem:schur_complement}) we have $\begin{pmatrix}
	A & B \\
	B^* & C
	\end{pmatrix} \geq 0 \iff C \geq 0$, $(\id - C C^{-1})B^* = 0$ and $A \geq B C^{-1}B^*$. Applying this identity to the second dual form~\eqref{eq:dual2_main} we find the optimal choice for the $A_i$ variables is $A_i = C_{i} C_{i+1}^{-1} C_{i}$ for $1 \leq i \leq k-1$ and $A_k = C_{k}\sigma^{-1}C_{k}$. For this particular choice the objective function becomes 
	\begin{equation}
	\sum_{i=1}^{k-1} \frac{2^{k-i}}{2^k-1} \tr{C_{i}^2 C_{i+1}^{-1}} + \frac{1}{2^k-1} \tr{C_{k}^2 \sigma^{-1}}.
	\end{equation}
	This expression is a convex combination of terms of the form $\petQ{2}(A\|B) = \tr{A^2 B^{-1}}$, i.e. the Petz generalized mean of order $2$. We see for $\alpha_k = 2$ the problem reduces to
	\begin{equation}
	\begin{aligned}
	\min_{C_1}& \quad\tr{C_1^2 \sigma^{-1}} \\
	\mathrm{s.t.}& \quad C_1 \geq \rho\, .
	\end{aligned}
	\end{equation}
	For the feasible point $C_1 = \rho$ we recover $\petQ{2}(\rho \| \sigma) = \tr{\rho^2 \sigma^{-1}}$ and so $\petQ{2}(\rho \| \sigma) \geq \imQ{2}(\rho \| \sigma)$ and therefore by monotonicity of the logarithm $\petD{2}(\rho \|\sigma) \geq \imD{2}(\rho \| \sigma)$. Furthermore, if we drop the constraint $V_1 + V_1^* \geq 0$ from the definition of $\imD{2}(\rho \| \sigma)$ then one can show that $\imD{2}(\rho\| \sigma) = \petD{2}(\rho \| \sigma)$.
\end{remark}

Recall that given a $\rho \in \D(AB)$ and a R\'enyi divergence $\mathbb{D}$ we may define the conditional entropy as $\mathbb{H}^{\downarrow}(A|B) = - \mathbb{D}(\rho_{AB} \| \id_A \otimes \rho_B)$ and its optimized version as $\mathbb{H}^{\uparrow}(A|B) = \sup_{\sigma \in \D(B)} - \mathbb{D}(\rho_{AB} \| \id_A \otimes \sigma_B)$. The following proposition gives an explicit characterization of $\mathbb{H}^{\uparrow}$ for the iterated mean divergences.
\begin{proposition}\label{prop:imHup}
	Let $\rho \in \D(AB)$. Then 
	\begin{equation}
	\imHup{\alpha_k}(A|B)_{\rho} = \frac{1}{1-\alpha_k} \log \imQup{\alpha_k}(\rho)
	\end{equation}
	where
	\begin{equation}\label{eq:imQup}
	\begin{aligned}
	\imQup{\alpha_{k}}(\rho) = \max_{V_1, \dots, V_k, Z}& \,\, \left(\tr{\rho \frac{(V_1 + V_1^{*})}{2}} \right)^{\alpha_k}  \\
	\mathrm{s.t.}& \quad  \lambda_{\max}(\ptr{A}{Z}) = 1 \\
	& \quad V_1 + V_1^* \geq 0 \\
	& \quad \begin{pmatrix} \id & V_1 \\ V_1^* & \frac{(V_2 + V_2^*)}{2} \end{pmatrix} \geq 0 \quad \begin{pmatrix} \id & V_2 \\ V_2^* & \frac{(V_3 + V_3^*)}{2} \end{pmatrix} \geq 0 \quad \cdots \quad \begin{pmatrix} \id & V_k \\ V_k^{*} & Z \end{pmatrix} \geq 0 \ .
	\end{aligned}
	\end{equation}
	or equivalently
	\begin{equation}\label{eq:imQup-noZ}
	\begin{aligned}
	\imQup{\alpha_{k}}(\rho) = \max_{V_1, \dots, V_k}& \,\, \left(\tr{\rho \frac{(V_1 + V_1^{*})}{2}} \right)^{\alpha_k}  \\
	\mathrm{s.t.}& \quad \ptr{A}{V_k^* V_k} \leq \id_B \\
	& \quad V_1 + V_1^* \geq 0 \\
	& \quad \begin{pmatrix} \id & V_1 \\ V_1^* & \frac{(V_2 + V_2^*)}{2} \end{pmatrix} \geq 0 \quad \begin{pmatrix} \id & V_2 \\ V_2^* & \frac{(V_3 + V_3^*)}{2} \end{pmatrix} \geq 0 \quad \cdots \quad \begin{pmatrix} \id & V_{k-1} \\ V_{k-1}^{*} & \frac{(V_k + V_k^*)}{2} \end{pmatrix} \geq 0 \ .
	\end{aligned}
	\end{equation}
	\begin{proof}
		By the definition of $\imHup{\alpha_k}(A|B)$ we have
		\begin{align*}
		\imHup{\alpha_k}(A|B) &= \sup_{\sigma_B} - \imD{\alpha_{k}}(\rho_{AB}\| \id_A \otimes \sigma_B) \\
		&= \frac{1}{1-\alpha_{k}} \log \inf_{\sigma_B} \max_{V_1, \dots, V_k, Z} \,\,\alpha_k \tr{\rho \frac{(V_1 + V_1^{*})}{2}} - (\alpha_k - 1) \tr{(\id_A \otimes \sigma_B) Z} \\
		& \qquad \qquad \qquad\mathrm{s.t.} \quad\,\, V_1 + V_1^* \geq 0 \\
		& \qquad \qquad \qquad \quad \,\, \begin{pmatrix} \id & V_1 \\ V_1^* & \frac{(V_2 + V_2^*)}{2} \end{pmatrix} \geq 0 \quad \begin{pmatrix} \id & V_2 \\ V_2^* & \frac{(V_3 + V_3^*)}{2} \end{pmatrix} \geq 0 \quad \cdots \quad \begin{pmatrix} \id & V_k \\ V_k^{*} & Z \end{pmatrix} \geq 0.
		\end{align*}
		Now consider the space 
		\begin{align*}
		\mathscr{M} &= \Big\{(V_1, \dots, V_k,Z) \in \Lin(AB)^{k+1} : \\
		 & \qquad V_1 + V_1^* \geq 0 \quad \begin{pmatrix} \id & V_1 \\ V_1^* & \frac{(V_2 + V_2^*)}{2} \end{pmatrix} \geq 0 \quad \begin{pmatrix} \id & V_2 \\ V_2^* & \frac{(V_3 + V_3^*)}{2} \end{pmatrix} \geq 0 \quad \cdots \quad \begin{pmatrix} \id & V_k \\ V_k^{*} & Z \end{pmatrix} \geq 0\Big\},
		\end{align*}
		and the function $f:\D(B) \times \mathscr{M} \rightarrow \RR$ defined as
		$$
		f(\sigma_B, V_1, \dots, V_k, Z) = \alpha_k \tr{\rho \frac{(V_1 + V_1^{*})}{2}} - (\alpha_k - 1) \tr{(\id_A \otimes \sigma_B) Z}.
		$$
		Note that $\mathscr{M}$ is a convex set, $\D(B)$ is a compact and convex set and $f$ is a continuous function. Additionally, $f$ is both convex and concave in each argument -- treating $(V_1, \dots, V_k, Z)$ as one argument. Now we have
		\begin{equation*}
		\begin{aligned}
		\inf_{\sigma_B}\max_{V_1, \dots, V_k, Z} f(\sigma_B, V_1, \dots, V_k, Z) &\geq \max_{V_1, \dots, V_k, Z}\inf_{\sigma_B} f(\sigma_B, V_1, \dots, V_k, Z) \\
		&= \max_{V_1, \dots, V_k, Z}\min_{\sigma_B} f(\sigma_B, V_1, \dots, V_k, Z) \\
		&= \min_{\sigma_B} \max_{V_1, \dots, V_k, Z} f(\sigma_B, V_1, \dots, V_k, Z) \\
		&\geq \inf_{\sigma_B}\max_{V_1, \dots, V_k, Z} f(\sigma_B, V_1, \dots, V_k, Z)
		\end{aligned}
		\end{equation*}
		where the second line follows from the fact that $\D(B)$ is compact and $f$ is continuous on $\D(B)$ and the third line from Sion's minimax theorem. Thus, we have 
		\begin{equation*}
		\inf_{\sigma_B}\max_{V_1, \dots, V_k, Z} f(\sigma_B, V_1, \dots, V_k, Z) = \max_{V_1, \dots, V_k, Z}\min_{\sigma_B} f(\sigma_B, V_1, \dots, V_k, Z)
		\end{equation*}
		and so we can interchange the $\inf \max$ in our optimization for a $\max \min$. 
		Now as $\max_{\sigma_B} \tr{(\id_A\otimes \sigma_B) Z} = \lambda_{\max}(\ptr{A}{Z})$ we can write
		\begin{align*}
		\imHup{\alpha_k}(A|B) &= \frac{1}{1-\alpha_{k}} \log \max_{V_1, \dots, V_k, Z} \,\,\alpha_k \tr{\rho \frac{(V_1 + V_1^{*})}{2}} - (\alpha_k - 1) \lambda_{\max}(\ptr{A}{Z}) \\
		& \qquad \qquad \qquad\mathrm{s.t.} \quad\,\, V_1 + V_1^* \geq 0 \\
		& \qquad \qquad \qquad \quad \,\, \begin{pmatrix} \id & V_1 \\ V_1^* & \frac{(V_2 + V_2^*)}{2} \end{pmatrix} \geq 0 \quad \begin{pmatrix} \id & V_2 \\ V_2^* & \frac{(V_3 + V_3^*)}{2} \end{pmatrix} \geq 0 \quad \cdots \quad \begin{pmatrix} \id & V_k \\ V_k^{*} & Z \end{pmatrix} \geq 0.
		\end{align*}
		Finally, by applying the same rescaling arguments used in the proof of property 1 in Proposition~\ref{prop:properties} we can homogenize the objective function to remove the second term and add the constraint $\lambda_{\max}(\ptr{A}{Z}) = 1$. After doing so we arrive at the first desired expression~\eqref{eq:imQup}. 
		
		To derive the second expression we first note that from Lemma~\ref{lem:schur_complement} it follows that the final positive-semidefinite constraint in~\eqref{eq:imQup} is equivalent to $Z \geq V_k^* V_k$. This condition, together with the fact that $V_k^* V_k \geq 0$, implies that $1 = \lambda_{\max}(\ptr{A}{Z}) \geq \lambda_{\max}(\ptr{A}{V_k^* V_k}) \geq 0$.  Now notice that for any feasible point $(V_1, \dots, V_k, Z)$ of~\eqref{eq:imQup}, the point $(V_1, \dots, V_k, \tfrac{V_k^*  V_k}{\lambda_{\max}(\ptr{A}{V_k^* V_k})})$ is also feasible and has the same objective value, we may therefore restrict our consideration to feasible points of this latter form. Furthermore, we have $\lambda_{\max}(\ptr{A}{V_k^* V_k}) \leq 1 \iff \ptr{A}{V_k^* V_k} \leq \id_B$. We now have a bijection between feasible points $(V_1, \dots, V_k)$ of~\eqref{eq:imQup-noZ} and feasible points $(V_1, \dots, V_k, \tfrac{V_k^*  V_k}{\lambda_{\max}(\ptr{A}{V_k^* V_k})})$ of~\eqref{eq:imQup} which preserves objective values and therefore the two optimizations are equivalent.
	\end{proof}
\end{proposition}

\begin{remark}[Relation to $H_{\min}(A|B)$]\label{rem:H2vsHmin}
	In Proposition~\ref{prop:properties} (see~\eqref{eq:property-ordering}) it was shown via an application of the Cauchy-Schwarz inequality that $\imD{\alpha_{k}}(\rho\|\sigma) \leq \imD{\alpha_{k-1}}(\rho\|\sigma)$, which in turn implies $\imHup{\alpha_{k}}(A|B) \geq \imHup{\alpha_{k-1}}(A|B)$. Applying the Cauchy-Schwarz inequality to the objective function of $\imHup{2}(A|B)$ we see that
	\begin{align*}
	- 2 \log \max_{V_1} \tr{\rho(V_1 + V_1^*)/2} &\leq -2 \log \max_{V_1} \tr{\rho V_1^* V_1}^{1/2} \\
	&= -\log \max_{V_1} \tr{\rho V_1^* V_1}.
	\end{align*}
	Therefore we have 
	\begin{align*}
	\imHup{2}(A|B) \geq -\log \max& \quad \tr{\rho V_1^* V_1} \\
	\mathrm{s.t.}& \quad \ptr{A}{V_1^* V_1} \leq \id_B \\
	& \quad V_1^* + V_1 \geq 0. 
	\end{align*}
	Let us compare this optimization with the min-entropy
	\begin{align*}
	H_{\min}(A|B) = -\log \max_{M \geq 0}& \quad \tr{\rho M} \\
	\mathrm{s.t.}& \quad \ptr{A}{M} \leq \id_B.
	\end{align*}
	As $V_1^*V_1 \geq 0$ we see that for each feasible point $V_1$ of the first optimization $V_1^* V_1$ is a feasible point of the second optimization with the same objective value. Conversely, for any feasible point $M$ of the second optimization, $V_1 = M^{1/2}$ is a feasible point of the first optimization with the same objective value and so $\imHup{2}(A|B) \geq H_{\min}(A|B)$. Thus, the sequence of conditional entropies $\imHup{\alpha_k}(A|B)$ are each separated by a Cauchy-Schwarz inequality and first term $\imHup{2}(A|B)$ is separated by another application of the Cauchy-Schwarz inequality from $H_{\min}(A|B)$.
\end{remark}
\section{Application to device-independent cryptography}


In the following we consider the setup wherein there are two devices\footnote{We restrict to bipartite setting for simplicity but our techniques readily extend to multipartite settings.} (which we refer to as Alice and Bob) that receive inputs $X$ and $Y$ from some finite alphabets $\cX$ and $\cY$ and produce outputs $A$ and $B$ in some finite alphabets $\cA$ and $\cB$ respectively. During a single interaction we assume that the devices operate in the following way. A bipartite quantum state $\rho_{Q_AQ_B} \in \D(Q_A Q_B)$ is shared between the two devices and in response to the  inputs $x \in \cX, y \in \cY$ the devices perform the POVMs $\{M_{a|x}\}_{a \in \cA}, \{N_{b|y}\}_{b \in \cB}$ on their respective systems. Inputs are chosen according to some fixed distribution $\mu : \cX \times \cY \rightarrow [0,1]$ that is known to all parties. The conditional probability distribution that describes the input-output behaviour of the two devices is then given by
\begin{equation}\label{eq:conditional-dist}
p(a,b|x,y) = \tr{\rho_{Q_AQ_B} (M_{a|x} \otimes N_{b|y})}.
\end{equation}
In addition, we allow for the presence of an adversarial party (Eve) who holds a purification of the quantum state initially shared between Alice and Bob, i.e., there is some pure quantum state $\outer{\psi} \in \D(Q_A Q_B E)$ such that $\ptr{E}{\outer{\psi}} = \rho_{Q_AQ_B}$. Formally, this setting may be characterized by a tuple $(Q_A,Q_B,E,\ket{\psi}, \{M_{a|x}\}, \{N_{b|y}\})$ which we shall refer to as a \emph{strategy}.

Let $\cC$ be another finite alphabet and let $C: \cA \times \cB \times \cX \times \cY \rightarrow \cC$ be some function -- this function will act as a statistical test on the devices. Given a probability distribution $q : \cC \rightarrow [0,1]$ we say that a conditional distribution $p_{AB|XY}$ is \emph{compatible} with $q$ if for all $c \in \cC$ we have
\begin{equation}
\sum_{abxy: C(a,b,x,y) = c} \mu(x,y)p(a,b|x,y) = q(c).
\end{equation}
More generally we say that a strategy $S$ is compatible with the statistics $q$ if the conditional distribution induced by the strategy (cf.~\eqref{eq:conditional-dist}) is compatible with $q$. For a given statistical test $C$ we denote the collection of all strategies that are compatible with the statistics $q$ by $\strat_C(q)$. The \emph{post-measurement state} of a strategy $S = (Q_A,Q_B,E,\ket{\psi}, \{M_{a|x}\}, \{N_{b|y}\})$ is 
\begin{equation}
\rho_{ABXYE} = \sum_{abxy} \mu(x,y) \outer{abxy} \otimes \rho_E^{abxy}
\end{equation}
where
\begin{equation}
\rho_{E}^{abxy} = \ptr{Q_AQ_B}{(M_{a|x} \otimes N_{b|y} \otimes \id_E)\outer{\psi}}.
\end{equation}

Let $\mathcal{P}(\cC)$ denote the set of all probability distributions on the alphabet $\cC$. A \emph{global tradeoff function} for the statistical test $C$ is a function $f: \mathcal{P}(\cC) \rightarrow \RR$ such that 
\begin{equation}\label{eq:tradeoff_global}
f(q) \leq \inf_{\strat_C(q)} H(AB|XYE), 
\end{equation}
where the infimum is taken over post-measurement states of all strategies that are compatible with the statistics $q$. Similarly we say a function $f: \mathcal{P}(\cC) \rightarrow \RR$ is a \emph{local tradeoff function} for the statistical test $C$ if it satisfies 
\begin{equation}\label{eq:tradeoff_local}
f(q) \leq \inf_{\strat_C(q)} H(A|XE).
\end{equation} 

We shall now demonstrate how to compute device-independent lower bounds on~\eqref{eq:tradeoff_global} and~\eqref{eq:tradeoff_local} using the conditional entropies $\imHup{\alpha_k}(AB|XYE)$. Furthermore, by replicating the tradeoff function constructions presented in~\cite{BRC} for the min-entropy, we can also derive explicit affine tradeoff functions from the results of our optimizations. Therefore, the present analysis can be readily extended to a full security proof of a device-independent protocol through an application of the entropy accumulation theorem~\cite{DF,DFR}.

In order to compute lower bounds on the von Neumann entropy we note that by Proposition~\ref{prop:properties} we have $\swHup{\alpha_k}(AB|XYE) \geq \imHup{\alpha_k}(AB|XYE)$ for all $k \in \NN$ and therefore we also have $H(AB|XYE) \geq \imHup{\alpha_k}(AB|XYE)$. Hence it suffices to demonstrate device-independent lower bounds on the latter quantity. The following lemma rewrites $\imHup{\alpha_k}(AB|XYE)$ into a form which can then be relaxed to a semidefinite program via the NPA hierarchy.

\begin{lemma}\label{lem:Hi_rewritten}
	Let $\outer{\psi} \in \D(Q_A E)$, $\{M_a\}_{a \in \cA}$ be a POVM on $Q_A$ and  $\rho_{AE} = \sum_{a} \outer{a} \otimes \rho_{E}(a)$ be a cq-state where $\rho_E(a) = \ptr{Q_A}{(M_{a} \otimes \id) \outer{\psi}}$. Then, for each $k \in \NN$ we have 
	\begin{equation}
	\imHup{\alpha_k}(A|E) = \frac{\alpha_k}{1-\alpha_k} \log \imQdi{\alpha_k}
	\end{equation}
	where 
	\begin{equation}\label{eq:r_k}
	\begin{aligned}
	\imQdi{\alpha_k} = \max_{V_{j,a} : 1 \leq j \leq k, a \in \cA}& \quad \sum_a \tr{\left(M_a \otimes \frac{V_{1,a} + V_{1,a}^*}{2}\right) \outer{\psi}} \\
	\mathrm{s.t.}& \quad  \sum_{a} V_{k,a}^* V_{k,a} \leq \id_E \\
	& \quad V_{1,a} + V_{1,a}^* \geq 0 \qquad \qquad \qquad \,\, \text{for all }a \in \cA \\
	& \quad 2 V_{i,a}^* V_{i,a}  \leq V_{i+1,a} + V_{i+1, a}^* \qquad \text{for all }1\leq i \leq k-1\text{ and }a \in \cA 
	\end{aligned}
	\end{equation}
	\begin{proof}
		From Proposition~\ref{prop:imHup} we know that
		\begin{align*}
		\imHup{\alpha_{k}}(A|E) &= \frac{\alpha_k}{1-\alpha_k} \log \max_{V_1, \dots, V_k} \tr{\rho_{AE} \frac{(V_1 + V_1^{*})}{2}}  \\
		& \quad \ptr{A}{V_k^* V_k} \leq \id_B  \\
		& \quad V_1 + V_1^* \geq 0 \\
		& \quad \begin{pmatrix} \id & V_1 \\ V_1^* & \frac{(V_2 + V_2^*)}{2} \end{pmatrix} \geq 0 \quad \begin{pmatrix} \id & V_2 \\ V_2^* & \frac{(V_3 + V_3^*)}{2} \end{pmatrix} \geq 0 \qquad \cdots \qquad \begin{pmatrix} \id & V_{k-1} \\ V_{k-1}^* & \frac{(V_k + V_k^*)}{2} \end{pmatrix} \geq 0 \ .
		\end{align*}
		For $1 \leq i \leq k$ let $V_i = \sum_{a,b} \outer{a}{b} \otimes \hat{V}_{i}(a,b)$ for some $\hat{V}_i(a,b) \in \Lin(E)$. Taking the partial trace over $A$ in the objective function we can rewrite it as
		\begin{align*}
		\tr{\frac{V_1 + V_1^{*}}{2}\rho_{AE}} &= \sum_{a} \tr{\frac{\hat{V}_1(a,a) + \hat{V}_1^{*}(a,a)}{2} \rho_E(a)} \\
		&= \sum_{a} \tr{\frac{\hat{V}_1(a,a) + \hat{V}_1^{*}(a,a)}{2} \ptr{Q_A}{(M_a \otimes \id) \outer{\psi}} } \\
		&= \sum_{a} \tr{\ptr{Q_A}{\left(M_a \otimes \frac{\hat{V}_1(a,a) + \hat{V}_1^{*}(a,a)}{2} \right) \outer{\psi}} } \\
		&= \sum_{a} \tr{\left(M_a \otimes \frac{\hat{V}_1(a,a) + \hat{V}_1^{*}(a,a)}{2} \right) \outer{\psi}}.
		\end{align*}
		Now for a linear operator $X = \sum_{a,b} \outer{a}{b} \otimes X(a,b)$ acting on $AE$ consider the pinching map defined by the action $\mathcal{P}(X) = \sum_{a} \outer{a} \otimes X(a,a)$ that pinches in the classical basis of $A$ defined by the cq-state $\rho_{AE}$. Note that $\mathcal{P}$ is both CP and unital and so it preserves the semidefinite constraints, i.e., 
		\begin{equation*}
		\begin{pmatrix} \id & V_i \\ V_i^* & \frac{(V_{i+1} + V_{i+1}^*)}{2} \end{pmatrix} \geq 0  \implies \begin{pmatrix} \id & \mathcal{P}(V_i) \\ \mathcal{P}(V_i)^* & \frac{(\mathcal{P}(V_{i+1}) + \mathcal{P}(V_{i+1})^*)}{2} \end{pmatrix} \geq 0
		\end{equation*}
		and $V_1 + V_1^* \geq 0 \implies \cP(V_1) + \cP(V_1)^* \geq 0$. 
		Furthermore, the variable $W_k = \mathcal{P}(V_k)$ also satisfies the constraint $\ptr{A}{W_k^*W_k} \leq \id$ as 
		\begin{equation*}
		\begin{aligned}
		\ptr{A}{W_k^*W_k} &= \sum_a \hat{V}_{k}^*(a,a) \hat{V}_k(a,a) \\
		&\leq \sum_{a,b} \hat{V}_{k}^*(a,b) \hat{V}_k(a,b) \\
		&= \ptr{A}{V_k^* V_k} 	\\
		&\leq \id.
		\end{aligned}
		\end{equation*}
		Finally, the objective function is invariant under the pinching as it only contains block diagonal elements $\hat{V}_k(a,a)$. As such, for any feasible point of the optimization problem we can replace the variables with their respective pinchings to obtain another feasible point with the same objective function value. We may therefore restrict all variables in the optimization to take the form $V_i = \sum_{a} \outer{a} \otimes \hat{V}_i(a,a)$. Applying Lemma~\ref{lem:schur_complement} to the remaining block positive semidefinite constraints and relabelling $\hat{V}_i(a,a)$ to $V_{i,a}$, the result follows.
	\end{proof}
\end{lemma}

\begin{example}
	For the post-measurement state of a strategy $S = (Q_A,Q_B,E,\ket{\psi}, \{M_{a|x}\}, \{N_{b|y}\})$ we have
	\begin{equation}
	\begin{aligned}
	\imHup{2}(AB|X=x,Y=y,E) = -2 \log \max_{V_{ab} : a \in \cA, b \in \cB}& \quad \sum_{ab} \tr{\left(M_{a|x} \otimes N_{b|y} \otimes \frac{V_{ab} + V_{ab}^*}{2}\right) \outer{\psi}} \\ 
	\mathrm{s.t.}& \quad  \sum_{a} V_{ab}^* V_{ab} \leq \id_E \\
	& \quad V_{ab} + V_{ab}^* \geq 0
	\end{aligned}.
	\end{equation}
	This should be compared with the analogous optimization for the conditional min-entropy, i.e.,
	\begin{equation}
	\begin{aligned}
	H_{\min}(AB|X=x,Y=y,E) = - \log \max_{W_{ab}: a \in \cA, b \in \cB}& \quad \sum_{ab} \tr{\left(M_{a|x} \otimes N_{b|y} \otimes W_{ab}\right) \outer{\psi}} \\ 
	\mathrm{s.t.}& \quad  \sum_{a} W_{ab} \leq \id_E \\
	&\quad W_{ab} \geq 0
	\end{aligned}.
	\end{equation}
\end{example}

The rewriting of $\imHup{\alpha_k}(A|E)$ in Lemma~\ref{lem:Hi_rewritten} still refers to an explicit pair of Hilbert spaces $Q_A, E$ and an explicit state $\ket{\psi} \in Q_A E$. In order to compute device-independent lower bounds on the various entropic quantities we also take the supremum in \eqref{eq:r_k} over all pairs of Hilbert spaces, and all operators and states on those Hilbert spaces. As mentioned previously, in order to approximate this extended optimization in an efficient manner it is possible to relax the optimization problem to a semidefinite program using the NPA hierarchy~\cite{NPA_general}. Indeed, we can optimize over moment matrices generated by the monomials $\{\id\}\cup\{M_a\}_{a \in \cA} \cup \{V_{i,a}, V_{i,a}^*\}_{1 \leq i \leq k, a \in \cA}$. The operator inequalities can be replaced by localizing moment matrices and we can enforce that all $[M_a, V_{i,a'}] = 0$ for all $a,a' \in \cA$ and $1\leq i \leq k$. We are also free to impose statistical constraints on our devices, e.g., a Bell-inequality violation.\footnote{Such additional constraints are necessary in order to obtain non-zero values.}

\begin{remark}\label{rem:additional_cons}
	When $k=1$ (i.e. $\alpha_k = 2$) and we are optimizing in the device-independent setting we may impose some additional constraints on the operators $\{V_{1,a}\}_a$. Namely, we may assume that for all $a,b \in \cA$,
	\begin{equation}\label{eq:additional_cons}
	V_{1,a} V_{1,b}^* = \delta_{ab}\id.
	\end{equation}
	This allows us to remove certain monomials from the moment matrix of the relaxed problem, which makes the size of the SDP smaller. Moreover, this implies that the operators $V_{1,a}^* V_{1,a}$ form a set of orthogonal projections. As was shown in Remark~\ref{rem:H2vsHmin}, we can recover $H_{\min}(A|E)$ from $\imHup{2}(A|E)$ by an appropriate application of the Cauchy-Schwarz inequality. In that case the operators $\{V_{a}^*V_a\}_{a \in \cA}$ played the role of Eve's POVM $\{W_a\}_{a \in \cA}$. By adding the additional constraints~\eqref{eq:additional_cons} to the optimization problem defining $\imHup{2}(A|E)$ the operators $\{V_{a}^*V_a\}_{a \in \cA}$ now form a projective measurement. This can be an important additional constraint as imposing that measurements are projective when computing $H_{\min}(A|E)$ often speeds up its convergence in the NPA hierarchy. Thus, the constraints~\eqref{eq:additional_cons} can also be helpful in this regard. However, in order to impose these constraints we have to remove or relax the constraint $V_1 + V_1^* \geq 0$. In practice, when computing the rates in the proceeding sections, we remove the constraint $V_1 + V_1^* \geq 0$ as we did not observe any change as a result. It is also possible to include these additional constraints in the optimizations of $\imHup{\alpha_{k}}$ for $k\geq 2$. However, this requires additional considerations. We discuss this further in Appendix~\ref{app:additional_constraints}.
\end{remark}

A more detailed explanation of the SDP implementation is given in Appendix~\ref{app:implementation}. To help facilitate the use of our techniques we also provide a few coded examples~\cite{github_imdivergences}. The NPA hierarchy relaxations were computed using~\cite{ncpol2sdpa} and all SDPs were solved using the Mosek solver~\cite{mosek}. For simplicity we shall only consider the entropy of some fixed inputs $(X,Y)=(x_0,y_0)$ -- this reflects the scenario usually considered in device-independent protocols where certain inputs are dedicated to generating secret key or randomness. For this reason, in the applications section only, we will abuse notation and for a conditional entropy $\mathbb{H}$, we will write $\mathbb{H}(AB|E)$ and $\mathbb{H}(A|E)$ instead of $\mathbb{H}(AB|X=x_0,Y=y_0,E)$ and $\mathbb{H}(A|X=x_0,E)$ respectively where the choice of $x_0,y_0$ will be clear from the context or otherwise explicitly stated.

\subsection{Application: Randomness certification}\label{sec:application_re}

We applied the semidefinite relaxations of $\imQdi{\alpha_k}$ to compute device-independent lower bounds on $\imHup{4/3}(AB|E)$ and $\imHup{2}(AB|E)$ for different statistical constraints. Firstly, we considered the CHSH game which is defined by the function 
\begin{equation}
C_{\mathrm{CHSH}}(a,b,x,y) = \begin{cases}
1 \quad & \text{if } a\oplus b = xy \\
0 \quad & \text{otherwise.}
\end{cases}
\end{equation} 
In addition to this we also considered the situation where the devices are constrained by their full conditional distribution i.e., we record each input-output tuple as a separate score $C:(a,b,x,y) \mapsto (a,b,x,y)$. We compared these to a tight analytical bound on the local von Neumann entropy $H(A|X=0,E)$ which is known for the CHSH game~\cite{ARV,PABGMS}, numerical lower bounds on $H_{\min}$ and the recent numerical lower bounds on the von Neumann entropy which were developed in~\cite{TSGPL19} (we refer to these latter bounds as the TSGPL bounds). For both devices constrained by the CHSH game and devices constrained by their full conditional distribution we evaluate the entropy for the inputs $(x_0, y_0) = (0,0)$.

\begin{figure}
	\centering
	\definecolor{mycolor2}{rgb}{0.00000,0.44700,0.74100}%
	\definecolor{mycolor4}{rgb}{0.85000,0.32500,0.09800}%
	\definecolor{mycolor3}{rgb}{0.92900,0.69400,0.12500}%
	\definecolor{mycolor1}{rgb}{0.49400,0.18400,0.55600}%
	\begin{tikzpicture}
	
	\begin{axis}[%
	width=4in,
	height=2.8in,
	scale only axis,
	xmin=0.75,
	xmax=0.854,
	ymin=0,
	ymax=1.6,
	grid=major,
	xlabel={CHSH score},
	ylabel={Bits},
	xtick={0.75, 0.77, 0.79, 0.81, 0.83, 0.85},
	axis background/.style={fill=white},
	legend style={at={(0.4,0.95)},legend cell align=left, align=left, draw=white!15!black}
	]
	\addplot[smooth, color=mycolor1, line width=1.2pt] table[col sep=comma] {h43_global_chsh_tikz.dat};
	\addlegendentry{$\imHup{4/3}(AB|E)$}
	\addplot[smooth, color=mycolor2, line width=1.2pt] table[col sep=comma] {h2_global_chsh_tikz.dat};
	\addlegendentry{$\imHup{2}(AB|E)$}
	\addplot[smooth, color=mycolor3, line width=1.2pt] table[col sep=comma] {hmin_global_chsh_tikz.dat};
	\addlegendentry{$H_{\min}(AB|E)$}
	\addplot[smooth, color=mycolor4, line width=1.2pt] table[col sep=comma] {chsh_analytic_tikz.dat};
	\addlegendentry{$H(A|E)$ analytic}	
	\end{axis}
	\end{tikzpicture}%
	\caption{Comparison of lower bounds on $H(AB|E)$ for quantum devices that constrained to achieve some minimal CHSH score.}
	
	\label{fig:chsh_comparison}
\end{figure}
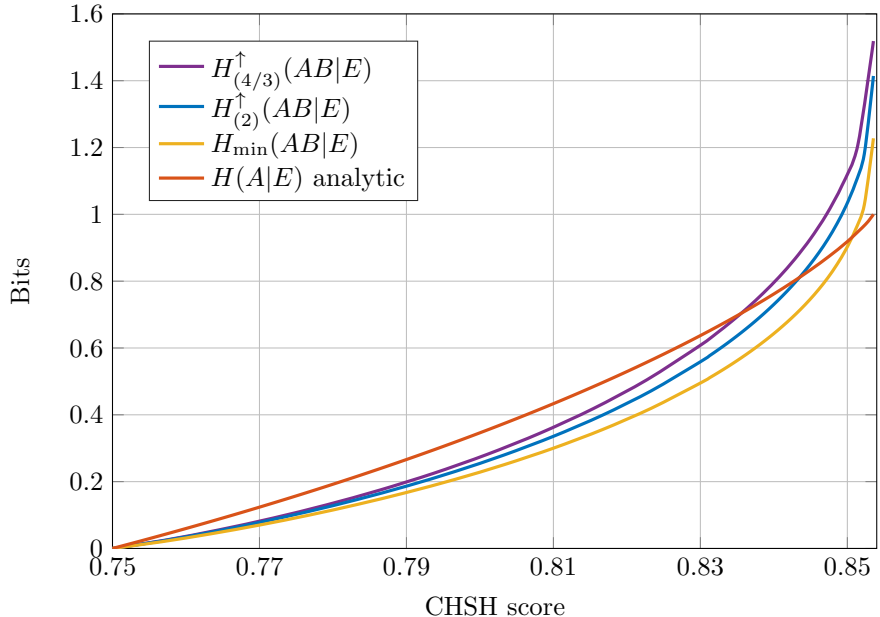

In Figure~\ref{fig:chsh_comparison} we plot lower bounds on the global entropies of Alice and Bob when their devices are constrained to achieve a minimal CHSH score. In the plot we observe a separation between the three curves that we compute numerically. That is, as we decrease $\alpha_k$ towards $1$ we see vsible improvements on the certifiable rates. For larger CHSH scores we observe that the lower bounds for both $\imHup{4/3}(AB|E)$ and $\imHup{2}(AB|E)$ can be used to certify substantially more randomness than $H_{\min}(AB|E)$. However, all three curves eventually drop below the randomness certified by the tight analytical bound on $H(A|E)$.

Recent device-independent experiments~\cite{diexperiment1,diexperiment2} have relied on measuring entangled photons in order to generate their nonlocal correlations. A major source of noise in these systems comes from inefficient detectors or losses during transmission of the photons. We model this noise by a single parameter $\eta \in [0,1]$ which characterizes the probability that after a photon has been produced by the source it is successfully transmitted and detected. For simplicity, we use the same $\eta$ for the photons of each party. In order to avoid a detection loophole in the experiment~\cite{Eberhard}, all failed detection events are recorded as the outcome $0$. This noise transforms the noiseless conditional probability distribution produced by the two parties in the following way 
\begin{equation}
p(a,b|x,y) \mapsto \eta^2 p(a,b|x,y) + \eta(1-\eta)(\delta_{a0} p(b|y) + \delta_{b0} p(a|x)) + \delta_{a0}\delta_{b0}(1-\eta)^2,
\end{equation}
where $\delta_{ij}$ is the Kronecker delta function. In order to generate valid quantum probability distributions we consider a two qubit setup with a state $\ket{\psi_\theta} = \cos(\theta) \ket{00} + \sin(\theta) \ket{11}$ with $\theta \in (0,\pi/4]$ and two-outcome qubit POVMs of the form $\{M, \id - M\}$ where $M = \outer{v}$ with $\ket{v} = \cos(\phi/2) \ket{0} + \sin(\phi/2) \ket{1}$ and $\phi \in (-\pi,\pi]$. We assume that $\cA = \cB = \cX = \cY = \{0,1\}$.

\begin{figure}[t]
	\centering
	\definecolor{mycolor2}{rgb}{0.00000,0.44700,0.74100}%
	\definecolor{mycolor4}{rgb}{0.85000,0.32500,0.09800}%
	\definecolor{mycolor3}{rgb}{0.92900,0.69400,0.12500}%
	\definecolor{mycolor1}{rgb}{0.49400,0.18400,0.55600}%
	\definecolor{mycolor5}{rgb}{0.4660, 0.6740, 0.1880}%
	\begin{tikzpicture}
	
	\begin{axis}[%
	width=4in,
	height=2.8in,
	scale only axis,
	xmin=0.7,
	xmax=1.0,
	ymin=0,
	ymax=2.0,
	grid=major,
	xlabel={Detection efficiency ($\eta$)},
	ylabel={Bits},
	xtick={0.7, 0.75, 0.8, 0.85, 0.9, 0.95, 1.0},
	axis background/.style={fill=white},
	legend style={at={(0.5,0.95)},legend cell align=left, align=left, draw=white!15!black}
	]
	\addplot[smooth, color=mycolor1, line width=1.2pt] table[col sep=comma] {h43_global_DE_l2_tikz.dat};
	\addlegendentry{$\imHup{4/3}(AB|E)$}
	\addplot[smooth, color=mycolor2, line width=1.2pt] table[col sep=comma] {h2_global_DE_l2_tikz.dat};
	\addlegendentry{$\imHup{2}(AB|E)$}
	\addplot[smooth, color=mycolor3, line width=1.2pt] table[col sep=comma] {hmin_global_DE_l2_tikz.dat};
	\addlegendentry{$H_{\min}(AB|E)$}
	\addplot[smooth, color=mycolor4, line width=1.2pt] table[col sep=comma] {chsh_analytic_de_tikz.dat};
	\addlegendentry{$H(A|E)$ analytic}
	\addplot[smooth, color=mycolor5, line width=1.2pt] table[col sep=comma] {ernest_global_de_tikz.dat};
	\addlegendentry{$H(AB|E)$ TSGPL bound}
	\end{axis}
	\end{tikzpicture}%
	\caption{Comparison of lower bounds on $H(AB|E)$ for quantum devices with inefficient detectors.}
	\label{fig:de_comparison}
\end{figure}
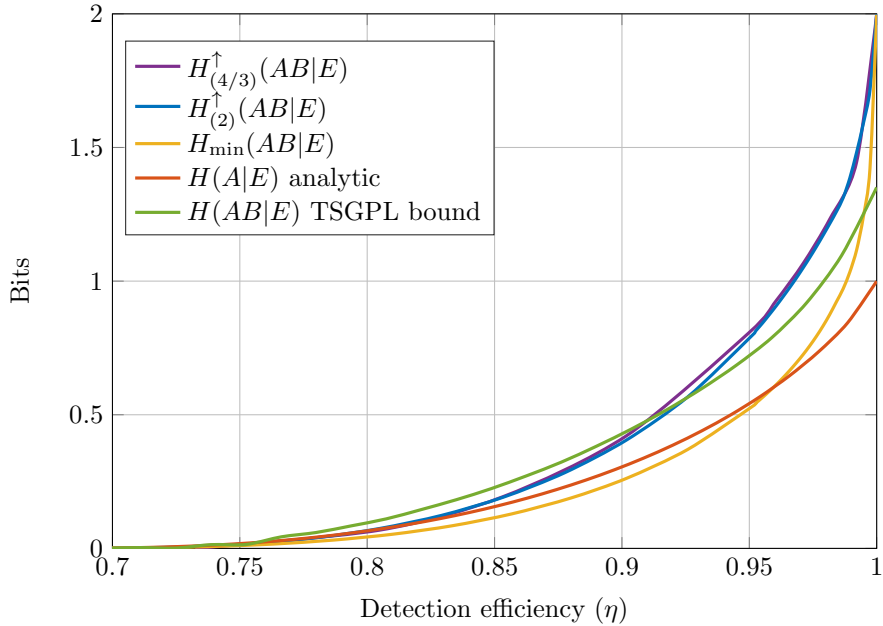

In Figure~\ref{fig:de_comparison} we compare lower bounds on the randomness certified by the different conditional entropies when the devices operate with inefficient detectors. In the SDPs we implement a constraint on the full conditional probability distribution of the devices, which is generated by some two-qubit model as described above. At each data point we also optimize the choice of two-qubit system in order to maximize the entropy bound using the iterative procedure described in~\cite{ATL16}.\footnote{This optimization is important when in the presence of inefficient detectors. For example, if we always use the two-qubit system which achieves Tsirelson's bound for the CHSH game in the noiseless case then we could not certify any entropy for detection efficiencies lower than $\eta \approx 0.83$. However, by allowing ourselves to optimize over partially entangled states we can certify entropy down to detection efficiencies of $\eta \approx 0.67$.} We compare the lower bounds produced by our SDPs again with the analytical bound of~\cite{ARV,PABGMS}, numerical bounds on $H_{\min}$ and the numerical techniques of~\cite{TSGPL19}. Note that the curves produced by the authors of~\cite{TSGPL19} also constrain the full conditional probability distribution of the devices. However, they did not implement the iterative optimization procedure, choosing instead to select a two-qubit system which maximized the CHSH score for a given detection efficiency. 

Observing the curves in the plot, we see that as before $\imHup{4/3}(AB|E)$ and $\imHup{2}(A|E)$ can be much larger than $H_{\min}(AB|E)$ and that the difference is more pronounced in this case. Moreover, by constraining the devices by the full conditional distribution we find a much larger improvement over the analytical bound on $H(A|E)$ which is only constrained by the CHSH game. Through our optimization over two-qubit systems we were also able to find two qubit systems that can certify the upper bound of two bits of randomness in the noiseless case. Unlike in Figure~\ref{fig:chsh_comparison} we find in this case a negligible difference between the randomness certified by $\imHup{4/3}(AB|E)$ and $\imHup{2}(AB|E)$. Comparing with the TSGPL bound we find that our optimized curves can certify more randomness in the lower noise regimes ($\eta > 0.92$). However for higher noise the TSGPL bound outperforms our method in this setting.

\subsection{Application: Quantum key distribution}

Continuing the comparison of entropy bounds for systems with inefficient detectors, we look at how this noise affects the rates of DI-QKD. Again we will consider devices that are constrained by the full conditional probability distribution, as was the case in Figure~\ref{fig:de_comparison}.  However, here we consider two separate setups. Firstly, we look at the $2$-input $2$-output setting, i.e., $\cA=\cB=\cX = \{0,1\}$ and $\cY = \{0,1,2\}$. We give Bob a third input which will act as his key-generation input, e.g., the key will be generated from the outputs of the devices on the input pair $(X,Y) = (0,2)$.\footnote{We still refer to this as the two-input two-output setting (despite Bob having a third input for key generation) as we only constrain the joint distribution of the devices on inputs $(x,y) \in \{0,1\}^2$.} Ideally, the correlations between Alice and Bob on this input pair are such that $H(A|B)$ is small. We generate the correlations of these devices with the same two-qubit model introduced in Section~\ref{sec:application_re}. As a novel comparison, we also look at the $2$-input $3$-output, i.e., $\cA=\cB = \{0,1,2\}$, $\cX=  \{0,1\}$ and $\cY = \{0,1,2\}$. We generate probability distributions for these devices using a two-qutrit model.\footnote{As before, we assume an explicit model for the devices in order to generate valid quantum conditional probability distributions. A parametrization also allows us to optimize the distribution in order to maximize the rates. However, the bounds on the rates are still device-independent as the SDP is only constrained by the conditional probability distribution and is not concerned with the model used to generate it.} For the two-qutrit state we consider the following family 
\begin{equation}
\sin(\theta) \cos(\phi) \ket{00} + \sin(\theta) \sin(\phi) \ket{11} + \cos(\theta) \ket{22},
\end{equation}
where $\theta \in [0, \pi]$ and $\phi \in [0, 2\pi)$. Furthermore, we assume that each measurement is a three outcome projective qutrit measurement and we use the parametrization given in~\cite{JFL16}. When considering no-click events, Alice and Bob both record these as the outcome $0$, except when Bob receives his key generating input $y=2$. When Bob inputs $y=2$ he no longer records a no-click as the outcome $0$ but rather as a new outcome $\perp$. This means that Bob has the potential to receive three or four outcomes whenever he inputs his key-generating input. Retaining this information allows us to reduce $H(A|B,X=0,Y=2)$ and further improve the key rate~\cite{ML12}.

\begin{figure}[t!]
	\centering
	\definecolor{mycolor2}{rgb}{0.00000,0.44700,0.74100}%
	\definecolor{mycolor4}{rgb}{0.85000,0.32500,0.09800}%
	\definecolor{mycolor3}{rgb}{0.92900,0.69400,0.12500}%
	\definecolor{mycolor1}{rgb}{0.49400,0.18400,0.55600}%
	\definecolor{mycolor5}{rgb}{0.4660, 0.6740, 0.1880}%
	\begin{tikzpicture}[spy using outlines={rectangle, magnification=3,connect spies}]
	
	\begin{axis}[%
	width=4in,
	height=2.8in,
	scale only axis,
	xmin=0.84,
	xmax=1.0,
	ymin=0,
	ymax=1.0,
	grid=major,
	xlabel={Detection efficiency ($\eta$)},
	ylabel={Key generation rate},
	xtick={0.84,0.86, 0.88, 0.9, 0.92, 0.94, 0.96, 0.98, 1},
	axis background/.style={fill=white},
	legend style={at={(0.55,0.95)},legend cell align=left, align=left, draw=white!15!black}
	]
	\addplot[ color=mycolor5, line width=1.pt] table[col sep=comma] {h87_de_optbin_tikz.dat};
	\addlegendentry{$\imHup{8/7}(A|E)- H(A|B)$}
	\addplot[smooth, color=mycolor1, line width=1.pt] table[col sep=comma] {h43_de_optbin_tikz.dat};
	\addlegendentry{$\imHup{4/3}(A|E)- H(A|B)$}
	\addplot[smooth, color=mycolor2, line width=1.pt] table[col sep=comma] {h2_de_optbin_tikz.dat};
	\addlegendentry{$\imHup{2}(A|E)- H(A|B)$}
	\addplot[smooth, color=mycolor3, line width=1.pt] table[col sep=comma] {hmin_de_optbin_tikz.dat};
	\addlegendentry{$H_{\min}(A|E)- H(A|B)$}
	\addplot[ color=mycolor4, line width=1.pt] table[col sep=comma] {h_rates_tikz.dat};
	\addlegendentry{$H(A|E) - H(A|B)$ analytic}
	\coordinate (insetPosition) at (axis cs:0.85,0.2);
	\coordinate (pos1big) at (axis cs:0.84,0.0);
	\coordinate (pos2big) at (axis cs:0.88,0.05);
	\coordinate (topleftbig) at (axis cs:0.84,0.05);
	\coordinate (bottomrightbig) at(axis cs:0.88,0.0);
	\end{axis}
	\begin{axis}[at={(insetPosition)},anchor={outer south west},
	width=2in,
	height=1.5in,
	xmin=0.84,
	xmax=0.88,
	ymin=0,
	ymax=0.05,
	xtick = {0.86},
	xticklabels={,,},
	ymajorticks=false,
	axis background/.style={fill=white}
	]
	\addplot[ color=mycolor5, line width=1.pt] table[col sep=comma] {h87_de_optbin_tikz.dat};
	\addplot[smooth, color=mycolor1, line width=1.2pt] table[col sep=comma] {h43_de_optbin_tikz.dat};
	\addplot[smooth, color=mycolor2, line width=1.pt] table[col sep=comma] {h2_de_optbin_tikz.dat};
	\addplot[ smooth, color=mycolor4, line width=1.2pt] table[col sep=comma] {h_rates_tikz.dat};
	\coordinate (topleftsmall) at (axis cs:0.84,0.05);
	\coordinate (bottomrightsmall) at (axis cs:0.88,0.0);
	\end{axis}
	\draw (pos1big) rectangle (pos2big); 
	\draw (topleftbig) -- (topleftsmall);
	\draw (bottomrightbig) -- (bottomrightsmall);
	\end{tikzpicture}%
	\caption{Comparison of lower bounds on the asymptotic device-independent key generation rates achievable with 2-output devices.}
	\label{fig:qkd_comparison_2out}
\end{figure}
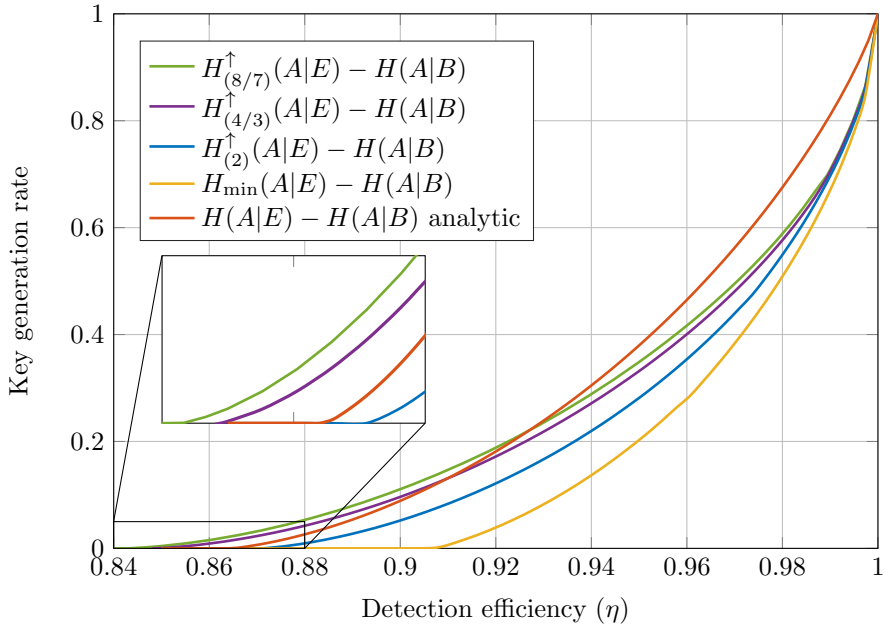

\begin{figure}[t!]
	\centering
	\definecolor{mycolor2}{rgb}{0.00000,0.44700,0.74100}%
	\definecolor{mycolor4}{rgb}{0.85000,0.32500,0.09800}%
	\definecolor{mycolor3}{rgb}{0.92900,0.69400,0.12500}%
	\definecolor{mycolor1}{rgb}{0.49400,0.18400,0.55600}%
	\definecolor{mycolor5}{rgb}{0.4660, 0.6740, 0.1880}%
	\begin{tikzpicture}[spy using outlines={rectangle, magnification=3,connect spies}]
	
	\begin{axis}[%
	width=4in,
	height=2.8in,
	scale only axis,
	xmin=0.84,
	xmax=1.0,
	ymin=0,
	ymax=1.6,
	grid=major,
	xlabel={Detection efficiency ($\eta$)},
	ylabel={Key generation rate},
	xtick={0.84, 0.86, 0.88, 0.9, 0.92, 0.94, 0.96, 0.98, 1},
	axis background/.style={fill=white},
	legend style={at={(0.55,0.95)},legend cell align=left, align=left, draw=white!15!black}
	]
	\addplot[smooth, color=mycolor1, line width=1.pt] table[col sep=comma] {h43_optbin_qutrits_tikz.dat};
	\addlegendentry{$\imHup{4/3}(A|E)- H(A|B)$}
	\addplot[smooth, color=mycolor2, line width=1.pt] table[col sep=comma] {h2_optbin_qutrits_tikz.dat};
	\addlegendentry{$\imHup{2}(A|E)- H(A|B)$}
	\addplot[smooth, color=mycolor3, line width=1.pt] table[col sep=comma] {hmin_optbin_qutrits_tikz.dat};
	\addlegendentry{$H_{\min}(A|E)- H(A|B)$}
	\addplot[ color=mycolor4, line width=1.pt] table[col sep=comma] {h_rates_tikz.dat};
	\addlegendentry{$H(A|E) - H(A|B)$ analytic}
	\coordinate (insetPosition) at (axis cs:0.85,0.2);
	\coordinate (pos1big) at (axis cs:0.84,0.0);
	\coordinate (pos2big) at (axis cs:0.88,0.05);
	\coordinate (topleftbig) at (axis cs:0.84,0.05);
	\coordinate (bottomrightbig) at(axis cs:0.88,0.0);
	\end{axis}
	\begin{axis}[at={(insetPosition)},anchor={outer south west},
	width=2in,
	height=1.5in,
	xmin=0.84,
	xmax=0.88,
	ymin=0,
	ymax=0.05,
	xtick = {0.86},
	xticklabels={,,},
	ymajorticks=false,
	axis background/.style={fill=white}
	]
	\addplot[smooth, color=mycolor1, line width=1.2pt] table[col sep=comma] {h43_optbin_qutrits_tikz.dat};
	\addplot[smooth, color=mycolor2, line width=1.2pt] table[col sep=comma] {h2_optbin_qutrits_tikz.dat};
	\addplot[ smooth, color=mycolor4, line width=1.2pt] table[col sep=comma] {h_rates_tikz.dat};
	\coordinate (topleftsmall) at (axis cs:0.84,0.05);
	\coordinate (bottomrightsmall) at (axis cs:0.88,0.0);
	\end{axis}
	\draw (pos1big) rectangle (pos2big); 
	\draw (topleftbig) -- (topleftsmall);
	\draw (bottomrightbig) -- (bottomrightsmall);
	\end{tikzpicture}%
	\caption{Comparison of lower bounds on the asymptotic device-independent key generation rates achievable with 3-output devices.}
	\label{fig:qkd_comparison_3out}
\end{figure}
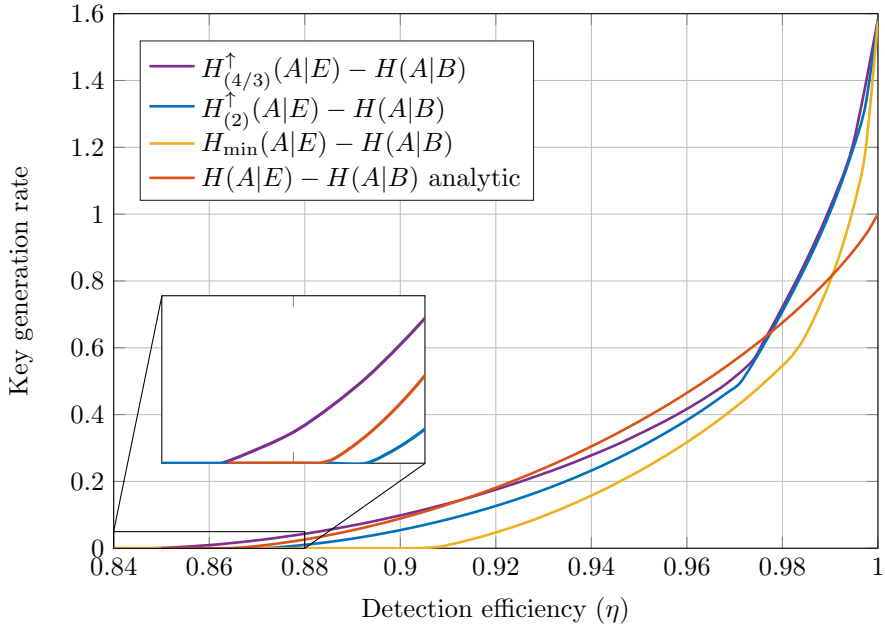

We consider a DI-QKD protocol with one-way error correction~\cite{ARV}. The asymptotic rate\footnote{Taking the asymptotic limit of finite round DI-QKD rates~\cite{ARV} and noting that the optimal one way error correction leaks approximately $n H(A|B)$ bits in an $n$ round protocol we recover the asymptotic i.i.d. rate~\cite{PABGMS}.} of such a protocol is given by the Devetak-Winter rate~\cite{DW05}
\begin{equation}
H(A|E) - H(A|B). 
\end{equation}
We apply our lower bounds on $H(A|E)$ to compute lower bounds on the asymptotic key rates. We compare our results again with the analytical bound on $H(A|E)$ and numerical bounds on $H_{\min}(A|E)$. The results for devices with two outputs are presented in Figure~\ref{fig:qkd_comparison_2out} and for devices with three outputs in Figure~\ref{fig:qkd_comparison_3out}. 

Producing high rates in DI-QKD is more difficult than just certifying randomness as the randomness needs to also be correlated between the two devices. In this application we see an even larger separation between the rates certified by the different entropies. In particular the minimal detection efficiency required to produce a positive rate differs substantially between the different entropies. For the curve generated by $H_{\min}$ we find the detection efficiency threshold is just below $0.91$, for $\imHup{2}$ it is just above $0.87$, for $\imHup{4/3}$ it is just below $0.85$ and for $\imHup{8/7}$ it is around 0.843. On the inset plot we zoom in on the region $[0.84, 0.88] \times [0.0, 0.05]$ and find that the detection efficiency threshold for the protocols based on $\imHup{8/7}$ and $\imHup{4/3}$ are significantly smaller than the protocol based on the CHSH game. Moreover the rates certified by $\imHup{8/7}$ are larger than those certified by the analytical bound on $H(A|E)$ for all $\eta < 0.92$. Similarly the rates certified by $\imHup{4/3}$ are larger than the rates certified by the analytical bound for all $\eta < 0.91$. 

For devices with two outputs, the rates are capped at one bit. However, for devices with three possible outputs we see in Figure~\ref{fig:qkd_comparison_3out} that it is possible to achieve a key rate of up to $-\log(1/3) \approx 1.59$ bits. For the curves based on $\imHup{4/3}(A|E)$ and $\imHup{2}(A|E)$ at around $\eta = 0.97$ and for the curve generated by $H_{\min}(A|E)$ at around $\eta = 0.985$ we see a sharp turn in the rates. This appears to correspond to a transition point where the optimal\footnote{This optima is not guaranteed to be a global optima.} state found by our optimization transitions from having a Schmidt rank of three to a Schmidt rank of two. Therefore, to the left of these points the optimal strategy found by the optimization could be implemented using a two qubit system and qubit POVMs. However, when moving from devices with two outputs to three outputs we do not see a significant change in the detection efficiency thresholds. For the curve generated by $H_{\min}$ the threshold is below $0.91$. For the curve generated by $\imHup{2}$ the threshold is around $0.87$ and for $\imHup{4/3}$ the threshold detection efficiency is again just below $0.85$. From our results it seems that the detection efficiency thresholds do not improve for two-qubit systems by moving from two-input two-outcome protocols to two-input three-outcome protocols. However, this may also be a consequence of the system optimization finding a local optima. Regardless our results show that we can implement DI-QKD protocols with two-qubit systems with smaller detection efficiencies. The achievability with two-qubit systems is of particular importance to experimental implementations where we seek robust protocols with simple setups.

Using the entropy accumulation theorem~\cite{DFR,DF} it would also be possible to calculate explicit lower bounds on the key rates for protocols with a finite number of rounds. In order to apply the EAT we must construct a min-tradeoff function (see~\eqref{eq:tradeoff_global} and~\eqref{eq:tradeoff_local}). By Lagrangian duality\footnote{We discuss in more detail how to construct min-tradeoff functions from our SDPs and the application to the EAT in Appendix~\ref{app:sdp2tradeoff}.} we can extract from the dual solution to our SDPs, an affine function
\begin{equation}
f : p_{AB|XY} \mapsto \alpha + \sum_{a,b,x,y} \lambda_{abxy} p(a,b|x,y).
\end{equation} 
For example, let us consider the two outcome protocols plotted in Figure~\ref{fig:qkd_comparison_2out}. For each curve and each value of $\eta$ we searched a two qubit system to generate a conditional distribution that maximized the rate. Let us take the two qubit system used for $\imHup{2}(A|E)$ at the point $\eta = 0.95$. This system is parameterized by six real numbers $(\theta, a_0, a_1, b_0, b_1, b_2)$. The state of the system is $\ket{\psi} = \cos(\theta) \ket{00} + \sin(\theta)\ket{11}$, Alice's measurements are defined by the projectors $ M_{0|x} = (\id + \cos(a_x) \sigma_z + \sin(a_x) \sigma_x)/2$ and Bob's measurements by the projectors $N_{0|y} = (\id + \cos(b_y) \sigma_z + \sin(b_y) \sigma_x )/2$. For this particular system the parameters were $(0.579, -0.161, 1.509, -1.207, 0.660, -0.177)$ and according to the solutions of the optimization we can use it to certify $0.415$ bits of entropy and a DI-QKD rate of $0.282$ bits when $\eta = 0.95$. Looking at the dual solution we can extract the function\footnote{For brevity we have only written the coefficients to three decimal places.  As such, this function can likely only guarantee a lower bound up to one or two decimal places.}
\begin{equation}
\begin{aligned}
g(p) := - 2 \log (& 93.340 - 1.558\, p(0,0|0,0) - 1.599\, p(0,0|0,1) + 93.940\, p(0,0|1,0) - 1.709 \, p(0,0|1,1) \\
+& 1.596\, p_A(0|0) -92.340\, p_B(0|0) - 92.334\, p_A(0|1) + 1.706\, p_B(0|1) \,)
\end{aligned}
\end{equation}
which should lower bound $\inf H(A|E)$. To obtain a min-tradeoff function we can take a first order Taylor expansion about some distribution, for example the distribution parametrizing the SDP, which gives us affine lower bounding function~\cite{BRC}.

\subsection{Application: Qubit randomness from sequential measurements}

As a final application we consider the question of how much local entropy can be device-independently certified from a two-qubit system. For example, it is well known that a score of $\cos(\pi/8)^2$ in the CHSH game self tests a maximally entangled two qubit state~\cite{PopescuRohrlich}. In such a case, the local statistics are uniformly distributed over $\{0,1\}$ and so this allows us to certify one bit of randomness using a two-qubit system. It has also been shown that up to two-bits of local randomness can be certified from a two-qubit system using strategies that include four-outcome qubit POVMs~\cite{APVW16}. 

It is also possible to consider scenarios wherein one party measures multiple times on their half of the two qubit system. By using \emph{unsharp} measurements~\cite{BLPY} it is possible to measure a two-qubit state such that the post-measurement state remains entangled. Therefore, a two-qubit state can be used to generate multiple instances of nonlocal correlations~\cite{SGGP} and in turn a sequence of certifiably random outcomes. The entropy of the sequence of measurement outcomes can then be lower bounded in a device-independent way by using an extension of the NPA hierarchy to sequential correlations~\cite{BBS}. In~\cite{BBS} the authors give an example~\cite[Section~4.1]{BBS} of a two-party scenario in which Bob measures his system twice. They gave an explicit two qubit setup, with a state $p \outer{\phi^+} + (1-p) \id /4$ where $ \ket{\phi^+} = \tfrac{1}{\sqrt{2}}(\ket{00} + \ket{11})$ and $p \in [0,1]$ such that $H_{\min}(B_1 B_2 |E) > 2$ for a range of $p$. Here $B_1$ refers to the outcome of Bob's first measurement and $B_2$ to his second. As before we look at the entropy only on particular inputs to the devices. 

\begin{figure}[t!]
	\centering
	\definecolor{mycolor2}{rgb}{0.00000,0.44700,0.74100}%
	\definecolor{mycolor4}{rgb}{0.85000,0.32500,0.09800}%
	\definecolor{mycolor3}{rgb}{0.92900,0.69400,0.12500}%
	\definecolor{mycolor1}{rgb}{0.49400,0.18400,0.55600}%
	\definecolor{mycolor5}{rgb}{0.4660, 0.6740, 0.1880}%
	\begin{tikzpicture}
	
	\begin{axis}[%
	width=4in,
	height=2.8in,
	scale only axis,
	xmin=0.0,
	xmax=0.1,
	ymin=0.0,
	ymax=2.5,
	grid=major,
	xlabel={Noise level ($p$)},
	ylabel={Bits},
	xtick={0.0, 0.02, 0.04, 0.06, 0.08, 0.1},
	axis background/.style={fill=white},
	legend style={at={(0.9,0.95)},legend cell align=left, align=left, draw=white!15!black}
	]
	\addplot[ color=mycolor2, line width=1.2pt] table[col sep=comma] {sequential_qubit_h2_tikz.dat};
	\addlegendentry{$\imHup{2}(B_1B_2|E)$}
	\addplot[ color=mycolor3, line width=1.2pt] table[col sep=comma] {sequential_qubit_hmin_tikz.dat};
	\addlegendentry{$H_{\min}(B_1B_2|E)$}
	\addplot[smooth, color=mycolor4, line width=1.2pt] table[col sep=comma] {nonsequential_qubit_hmin_tikz.dat};
	\addlegendentry{$H_{\min}(B|E)$ - CHSH strategy}
	\end{axis}
	\end{tikzpicture}%
	\caption{Lower bounds on the certifiable randomness produced by two sequential measurements on one half of the two-qubit state $p \outer{\phi^+} + (1-p) \id /4$.}
	\label{fig:sequential_comparison}
\end{figure}
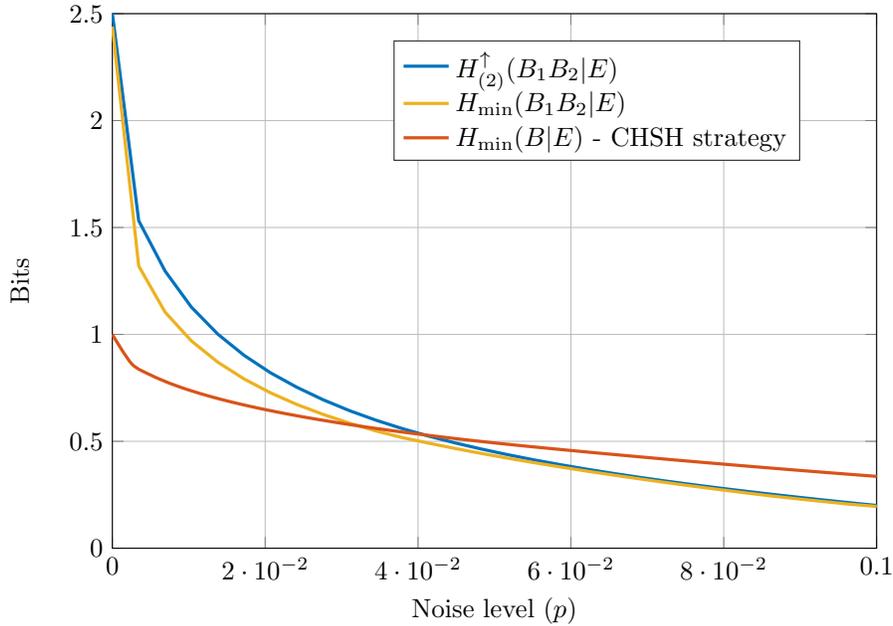

In Figure~\ref{fig:sequential_comparison} we reproduce Figure~3 from~\cite{BBS} which computes a lower bound on $H_{\min}(B_1B_2|E)$ and compares with the randomness certified by $H_{\min}(A|E)$ for a single two-outcome projective measurement. To illustrate our technique we also include a lower bound on $\imHup{2}(B_1B_2|E)$. We see that for low noise the randomness as measured by $\imHup{2}(B_1B_2|E)$ can be noticeably larger than $H_{\min}(B_1B_2|E)$. Unlike the previous two examples no concrete protocol or security proof was studied for this scenario and thus neither $\imHup{2}(B_1B_2|E)$ nor $H_{\min}(B_1B_2|E)$ correspond to actual rates.
However, the example does illustrate that our conditional entropies can also be easily computed in more exotic scenarios where previously bounds on $H_{\min}$ have been used. 

\section{Conclusion}
In this work we introduced a new family of R\'enyi divergences that correspond to convex optimization problems. We showed that the conditional entropies defined by these divergences are amenable to device-independent optimization and can be used as tools to derive numerical lower bounds on the conditional von Neumann entropy. We applied this to the task of computing lower bounds on the rates of device-independent randomness generation and quantum key distribution protocols. We compared the protocol rates derived from our techniques to the analytical bound of~\cite{PABGMS,ARV}, the numerical techniques of~\cite{TSGPL19} and bounds established via the min-entropy~\cite{KRS,NPS14,BSS14}. We found improvements over all three of these bounds in various settings.

In particular, when looking at randomness generation in low noise regimes we found improvements over all the previous methods. But in the higher noise regimes, our bounds typically were outperformed by the numerical techniques of~\cite{TSGPL19} in the scenarios where we could compare. 
However, this comparison has only been performed for some simple protocols where the data for~\cite{TSGPL19} is available. We suspect that our approach is more computationally efficient and thus could be used to analyze a wider range of scenarios. For example the computational efficiency allowed us to iteratively optimize over two-qubit protocols to improve the randomness certification rates up to the maximum of two bits. It is also possible that a combination of the two approaches could yield even higher rates. That is, our techniques could be used to search for optimized protocols and then if the TSGPL bound could be computed it may yield further improvements on the rates. 

When computing key rates for DI-QKD, we also looked at bipartite protocols using devices with two-outputs and three-outputs. There we found significant improvements in the minimal detection efficiency required to generate key without noisy preprocessing of the raw key. Moreover, in the regimes of higher noise all of these protocols were still implementable using entangled states of two qubits. It is possible that by further increasing the number of inputs/outputs or by searching for protocols compatible with higher dimensional systems that additional improvements could be made but we leave such an investigation to future work. Reducing the minimal detection efficiency is important for practical experiments, recent works~\cite{HSTRBS20,WAP20} have shown that noisy preprocessing of the raw key could also be used to improve minimal detection efficiency for a protocol based on the CHSH game. It would be interesting to see if this could be combined with our numerical techniques to further improve the detection efficiency threshold and design more robust device-independent protocols. 

We also demonstrated that min-tradeoff functions could be derived directly from solutions to our device-independent optimizations. These functions can be combined with the entropy accumulation theorem in order to construct simple security proofs for device-independent protocols~\cite{ARV,ADFRV}. Therefore, not only can our conditional entropies be used to derive lower bounds on the rates of various protocols but they can also be used directly with the EAT to establish their security proofs and compute finite key rates. We note that it is not clear if the TSGPL method can be used in the same way. 

As a final example, we also showed that our techniques could be used in conjunction with the newly introduced semidefinite hierarchy for sequentially generated quantum correlations~\cite{BBS}. Repeating an example from~\cite{BBS}, which looked at the randomness generated from two sequential measurements, we showed higher rate curves could be obtained by using $\imHup{2}$ as opposed to $H_{\min}$ which was the measure of randomness originally used in the example.

Several additional questions remain open from this work. Firstly, what can be said about the limit $\imD{\alpha_k}$ as $k \to \infty$? We know that it will be between the Umegaki divergence $D$ and the Belavkin-Staszewski divergence $\widehat{D}$, and we also know that it cannot always be equal to $\widehat{D}$ (there are some examples where already $\imD{2} < \widehat{D}$). If one can show that $\lim_{k \to \infty} \imD{\alpha_k} = D$, then this shows that our technique can approximate the conditional von Neumann entropy arbitrarily well. More generally, a very interesting question is whether other divergences that provide a tighter approximation to the Umegaki divergence (e.g., the sandwiched R\'enyi divergence) have a free variational expression as discussed in Remark~\ref{rem:dimension-free}.

Secondly, it would be interesting to see whether our computations can be made more efficient. For example, in Appendix~\ref{app:additional_constraints} we show that a particular dilation theorem can be applied to reduce the size of the optimization $\imHup{2}$ and speed up its convergence. It would be interesting to see whether one could extend this dilation theorem to other $\imHup{\alpha_k}$ to help improve their convergence and efficiency also. Additionally, it may be possible to reduce the size of the $\imHup{\alpha_k}$ optimizations by exploiting symmetries of the problem~\cite{R18} or by optimizing the choice of monomial sets generating the NPA moment matrices.

\section*{Acknowledgements}

This research is supported by the French National Research Agency via Project No. ANR-18-CE47-0011 (ACOM). PB thanks Ernest Tan for useful discussions and for providing data used in the comparisons with the TSGPL bound. PB also thanks Joseph Bowles for providing code used for generating the semidefinite relaxations in the sequential correlation example. 

\bibliographystyle{plain}
\bibliography{tradeoff}

\appendix
	
\section{SDP implementation details}\label{app:implementation}
\subsection{The NPA hierarchy}

In this subsection we briefly describe how we can use NPA hierarchy to optimize polynomials of bounded operators. For more details we refer the reader to the original paper~\cite{NPA_general}. Consider a Hilbert space $\cH$, a collection of bounded operators on $\cH$, $X =(X_1, \dots, X_n)$ and a state $\ket{\psi} \in \cH$. Call the elements in the collection $X$ \emph{letters}, then a \emph{word} consists of an arbitrary product of letters and their adjoints. The length of a word is the number of letters in the product. We consider $\id$ to be the empty word and define its length to be $0$. Let $\cW_k$ be the set of all words of length no larger than $k$. Now consider the matrix $\Gamma$ whose elements are indexed by words in the set $\cW$ and whose $(W_1, W_2)$ element corresponds to 
\begin{equation}
\Gamma_{(W_1, W_2)} = \tr{W_1^* W_2 \outer{\psi}}. 
\end{equation}  
It was shown in~\cite{NPA_general} that this matrix is PSD for all $k \in \NN$. We refer to such a matrix as a \emph{certificate} of level $k$.

Now suppose we are given a conditional probability distribution $p(a,b|x,y)$. We say $p$ has a \emph{quantum spatial realization} if there exists a Hilbert space $\cH$, a state $\ket{\psi} \in \cH$ and POVMs $\{M_{a|x}\}$, $\{N_{b|y}\}$ with $[M_{a|x}, N_{b|y}] = 0$ for all $(a,b,x,y)$ such that $p(a,b|x,y) = \tr{M_{a|x} N_{b|y} \outer{\psi}}$. The above construction allows us to derive necessary conditions for a distribution to have a quantum spatial realization. That is, we know if a quantum realization were to exist then for each $k \in \NN$ there exists a certificate of level $k$. Thus, we can look for a positive semidefinite matrix $\Gamma$ indexed by words on length no larger than $k$ generated from the set $\{\id\} \cup \{M_{a|x}\} \cup \{N_{b|y}\}$ which would be compatible with the distribution $p$. For example, we know constraints such as
$$
\Gamma_{(M_{a|x}, N_{b|y})} = \Gamma_{(N_{b|y}, M_{a|x})} = \Gamma_{(M_{a|x}N_{b|y}, \id)} = p(a,b|x,y)
$$
and 
$$
\Gamma_{(\id, \id)} = 1.
$$
After imposing all such constraints, finding a completion of the matrix that is positive semidefinite is an SDP and so can be computed efficiently. The authors of~\cite{NPA_general} also proved a converse statement: if for each $k \in \NN$ there exists a certificate of level $k$ then there exists a quantum realization of the probability distribution. 

This construction allows us to relax optimization problems of the form 
\begin{equation}\label{eq:poly_opt}
\max \tr{m(X) \outer{\psi}} 
\end{equation}
where $m(X)$ is some Hermitian polynomial of bounded operators and the maximization is taken over all Hilbert spaces $\cH$, all collections of bounded operators on that Hilbert space and all states $\ket{\psi} \in \cH$ to an SDP. We can add tracial constraints, e.g., $\tr{n(X) \outer{\psi}} = c$ for some polynomial $n(X)$, and also operator inequalities to the optimization~\eqref{eq:poly_opt}. Given a Hermitian polynomial $q(X) \geq 0$, if we have a quantum realization then the \emph{localizing matrix} $\Gamma^{\mathrm{loc}}$ indexed by words in $\cW_d$ whose entries are given by
\begin{equation}\label{eq:localizing_matrix}
\Gamma_{(W_1, W_2)}^{\mathrm{loc}} = \tr{W_1^* q(X) W_2 \outer{\psi}}
\end{equation}
is also PSD. Therefore, for each operator inequality we add to~\eqref{eq:poly_opt} we can relax the optimization by adding an additional localizing matrix.

\subsection{Further constraints for $\imHup{2}(A|E)$}\label{app:additional_constraints}

The following proposition, taken from~\cite{B84}, provides a dilation theorem which can be used to simplify our device-independent optimizations. 

\begin{proposition}[Proposition~1.~\cite{B84}]\label{prop:dilation}
	Let $n \in \NN$ and let $\{V_i : 1 \leq i \leq n\}$ be a collection of bounded linear operators on some Hilbert space $\cH$ such that $\sum_{i=1}^n V_i^* V_i \leq \id$. Then there exists a Hilbert space $\cK$, such that $\cH \subseteq \cK$, and a collection of bounded linear operators $\{S_i : 1 \leq i \leq n\}$ on $\cK$ satisfying
	\begin{enumerate}
		\item $S_i(\cH) \subseteq \cH$ for each $i \in \{1, \dots, n\}$.
		\item $S_i S_j^* = \delta_{ij} \id_{\cK}$ for each $i,j \in \{1, \dots, n\}$.
		\item $\sum_{i=1}^n S_i^*S_i \leq \id_{\cK}$. 
		\item $P_{\cH} S_i |_{\cH} = V_i$ for each $i \in \{1, \dots, n\}$.
	\end{enumerate}
	where $P_{\cH}$ is the projector onto the subspace $\cH$. 
\end{proposition}

The proof of the above proposition, see~\cite{B84}, gives a construction of the operators $S_i$. Briefly, it states that we find can some (possibly infinite-dimensional) Hilbert space $ \cL$ such that $\cK = \cH \oplus \cL$ and operators $S_i$ of the form 
\begin{equation}
S_i = \begin{pmatrix}
V_i & X_i \\
0 & Y_i 
\end{pmatrix}
\end{equation}
for some suitably chosen operators $X_i$ and $Y_i$. 

We now look to apply the this dilation theorem to improve convergence and efficiency of our device-independent optimizations of $\imHup{\alpha_k}$. Let us first describe how the above proposition can be used to improve the optimization of $\imHup{2}$, afterwards we shall describe the general case. Recall that $\inf \imHup{2}(A|E) = - 2 \log(Q_{(2)}^{\mathrm{DI}})$ where
\begin{equation}\label{eq:di_h2_optimization}
\begin{aligned}
Q_{(2)}^{\mathrm{DI}} = &\sup_{\{V_a\}_a, \{M_{a}\}_a, \outer{\psi}, Q_A \otimes E} \sum_a \tr{(M_a \otimes \frac{V_a + V_a^*}{2})\outer{\psi}}\\
&\qquad \mathrm{s.t.} \qquad \sum_a V_a^* V_a \leq \id_E \\
& \qquad \qquad V_a + V_a^* \geq 0 \qquad \text{for each }a \in \cA
\end{aligned}
\end{equation}
where the optimization is over all joint Hilbert spaces $Q_A E$, all states $\ket{\psi} \in Q_A E$, all POVMs $\{M_a\}_a$ on $Q_A$ and all collections of linear operators $V_a \in \Lin(E)$. For the moment we will drop the operator inequalities $V_a + V_a^* \geq 0$ from the optimization and later we shall discuss how to reinsert them.\footnote{By removing a constraint we still have a lower bound on the corresponding conditional entropy which is what is required by the device-independent applications.} In general this optimization would also be augmented with constraints on the local statistics generated by the POVMs $\{M_a\}$ and likely would also include a second system $Q_B$ with further POVMs. However, we deal with the simpler case here from which the general case follows readily. Furthermore, the SDP relaxations of this problem~\cite{NPA_general} provide lower bounds on the optimization even when the Hilbert spaces $Q_A$ and $E$ are infinite dimensional. 

Now consider a more restricted optimization
\begin{equation}\label{eq:di_h2_optimization_restricted}
\begin{aligned}
\widehat{Q}_{(2)}^{\mathrm{DI}} = &\sup_{\{S_a\}_a, \{M_{a}\}_a, \outer{\psi}, Q_A \otimes \widehat{E}} \sum_a \tr{(M_a \otimes \frac{S_a + S_a^*}{2})\outer{\psi}}\\
&\qquad \mathrm{s.t.} \qquad \sum_a S_a^* S_a \leq \id_{\widehat{E}} \\
& \qquad \qquad \qquad  S_a S_b^* = \delta_{ab} \id_{\widehat{E}} \quad \text{ for all }a,b\in \cA. 
\end{aligned}
\end{equation}
By Proposition~\ref{prop:dilation}, any feasible point of~\eqref{eq:di_h2_optimization} can be transformed into a feasible point of~\eqref{eq:di_h2_optimization_restricted} with the same objective value. Indeed, the proposition states that we can find a larger Hilbert space $\widehat{E} = E \oplus E^{\perp}$, with operators of the form $S_a = \begin{pmatrix}
V_a & X_a \\
0 & Y_a 
\end{pmatrix}$ satisfying the constraints of~\eqref{eq:di_h2_optimization_restricted}. Moreover, we can use an isometry $W: E\rightarrow \widehat{E}$ to embed the state $\ket{\psi} \in Q_A\otimes E$ in $Q_A \otimes \widehat{E}$, i.e. $W = \begin{pmatrix}
\id_E \\
0_{E^{\perp}}
\end{pmatrix}$. Defining $\outer{\widehat{\psi}} = (\id\otimes W) \outer{\psi} (\id \otimes W^*)$ we see that the objective value remains unchanged, 
\begin{equation}
\begin{aligned}
\sum_a \tr{(M_a \otimes \frac{S_a + S_a^*}{2})\outer{\widehat{\psi}}} &= 
\sum_a \tr{(M_a \otimes \frac{S_a + S_a^*}{2})(\id\otimes W) \outer{\psi} (\id \otimes W^*)} \\
&= \sum_a \tr{(M_a \otimes \frac{W^*S_aW + W^*S_a^*W}{2}) \outer{\psi}} \\
&= \sum_a \tr{(M_a \otimes \frac{V_a + V_a^*}{2}) \outer{\psi}}.
\end{aligned}
\end{equation}
Thus we have $\widehat{Q}_{(2)}^{\mathrm{DI}} \geq Q_{(2)}^{\mathrm{DI}}$. However, as the optimizations range over all Hilbert spaces (assuming also infinite dimensional) we have that any feasible point of~\eqref{eq:di_h2_optimization_restricted} is trivially a feasible point of~\eqref{eq:di_h2_optimization} and so $\widehat{Q}_{(2)}^{\mathrm{DI}} \leq Q_{(2)}^{\mathrm{DI}}$. Therefore we conclude that $\widehat{Q}_{(2)}^{\mathrm{DI}} = Q_{(2)}^{\mathrm{DI}}$ and we can impose the additional restrictions of~\eqref{eq:di_h2_optimization_restricted} when we drop the constraints $V_a + V_a^* \geq 0$. 

Unfortunately, the dilation theorem does not immediately apply to the optimization that includes the operator inequalities $V_a + V_a^* \geq 0$ as it need not hold that $S_a + S_a^* \geq 0$ if $V_a + V_a^* \geq 0$. One workaround is to drop these constraints from the optimization, which is was what was done when computing the rate plots from the main text. Alternatively, we can relax the constraint to a moment inequality as $\tr{(V_a + V_a^*) \outer{\psi}} \geq 0 \implies \tr{(S_a + S_a^*) \outer{\widehat{\psi}}} \geq 0$.

What remains is to consider how this dilation theorem may be used to impose additional constraints on the other conditional entropies $\imHup{\alpha_k}$. For simplicity, let us consider the case of $\alpha_k = 4/3$, for the other $\alpha_k$ the procedure remains the same. Recall that,
\begin{equation}\label{eq:di_h43_optimization}
\begin{aligned}
Q_{(4/3)}^{\mathrm{DI}} = &\sup_{\{V_{1,a}\}_a,\{V_{2,a}\}_a, \{M_{a}\}_a, \outer{\psi}, Q_A \otimes E} \sum_a \tr{(M_a \otimes \frac{V_{1,a} + V_{1,a^*}}{2})\outer{\psi}}\\
&\qquad \mathrm{s.t.} \qquad \sum_a V_{2,a}^* V_{2,a} \leq \id_E \\
&\qquad \qquad \qquad V_{1,a}^* V_{1,a} \leq \frac{V_{2,a} + V_{2,a}^*}{2} \quad \text{ for all }a\in\cA.
\end{aligned}
\end{equation}
Following the previous construction we can define a larger Hilbert space $\widehat{E}$ and some operators $\{S_{2,a}\}_a$ that play the role of $\{V_{2,a}\}$ but satisfy the additional restriction of being coisometries with orthogonal ranges. Unfortunately, we run into similar problems to the ones that we faced with the operator inequalities $V_a + V_a^* \geq 0$ when dilating $\imHup{2}$. If we embed $\{V_{1,a}\}$ and $\outer{\psi}$ using the isometry $W$ as before, the objective value remains unchanged but the constraints $ V_{1,a}^* V_{1,a} \leq \frac{V_{2,a} + V_{2,a}}{2}$ must be interpreted on the subspace $E$. This is because $ V_{1,a}^* V_{1,a} \leq \frac{V_{2,a} + V_{2,a}^*}{2} \centernot\implies W V_{1,a}^* V_{1,a}W^* \leq \frac{S_{2,a} + S_{2,a}^*}{2}$. To see this note that the left-hand-side of the second inequality has support only on the subspace $E$ but the right-hand-side may have support elsewhere and need not be positive semidefinite a priori.

Again, we can weaken this constraint from an operator inequality to a trace inequality 
\begin{equation}
\tr{V_{1,a}^* V_{1,a}\outer{\psi}} \leq \tr{\frac{V_{2,a} + V_{2,a}^*}{2} \outer{\psi}}.
\end{equation}
For this weaker constraint, its dilated counterpart $\tr{W V_{1,a}^* V_{1,a}W^* \outer{\widehat{\psi}}} \leq \tr{\frac{S_{2,a} + S_{2,a}^*}{2} \outer{\widehat{\psi}}}$ does hold true as $\tr{S_{2,a}\outer{\widehat{\psi}}} = \tr{V_{2,a} \outer{\psi}}$. However, after numerical testing we found that this weaker constraint often lead to much weaker results and so for all of the numerical examples we decided not to add any additional constraints to the optimizations of $\imHup{4/3}$. 

\subsection{Sufficient relaxation level to observe ordering}
We know for a given cq-state $\rho_{AE}$ that $\imHup{\alpha_k}(A|E) \geq \imHup{\alpha_{k-1}}(A|E) \geq H_{\min}(A|E)$. However, when we perform device-independent optimizations of these quantities we relax the optimization problem to a semidefinite program via the NPA hierarchy~\cite{NPA_general}. For a given level of relaxation, the corresponding relaxed problems need not always satisfy this ordering. However, it is possible to find a sufficient level of relaxation such that the ordering holds.

For example, consider the commuting operator version of the min-entropy problem
\begin{equation}\label{eq:hmin_noncommuting}
\begin{aligned}
- \log \max& \sum_a \tr{M_a W_a \outer{\psi}} \\
\mathrm{s.t.}& \quad \sum_a W_a \leq \id \\
& \quad W_a \geq 0 \qquad \text{for all } a \in \cA  \\
& \quad \sum_a M_a = \id \\
& \quad M_a \geq 0 \qquad \text{for all } a \in \cA \\
& \quad [M_a, W_b] = 0 \qquad \text{for all } a,b \in \cA
\end{aligned}
\end{equation}
and the corresponding problem for $\imHup{2}(A|E)$
\begin{equation}\label{eq:h2_noncommuting}
\begin{aligned}
- 2 \log \max& \sum_a \tr{M_a\frac{V_a + V_a^*}{2} \outer{\psi}} \\
\mathrm{s.t.}& \quad \sum_a V_a^*V_a \leq \id \\
& \quad V_a + V_a^* \geq 0 \qquad \text{for all } a \in \cA \\
& \quad \sum_a M_a = \id \\
& \quad M_a \geq 0  \qquad \text{for all } a \in \cA\\
& \quad [M_a, V_b^{(*)}] = 0 \qquad \text{for all } a,b \in \cA.
\end{aligned}
\end{equation}
By applying an appropriate Naimark dilation to the Hilbert space we may assume that $\{M_a\}$ forms a projective measurement.\footnote{We could also make this assumption for $\{W_a\}$. However, to then establish ordering we would have to include the additional constraints that were introduced in Appendix~\ref{app:additional_constraints}. For simplicity we do not consider this but the strategy for enforcing an ordering works in the same manner.}

We know from Remark~\ref{rem:H2vsHmin} that for an explicit state $\rho_{AE}$, $\imHup{2}(A|E)$ and $H_{\min}(A|E)$ are related by the Cauchy-Schwarz inequality
$$
\frac12\tr{M_a(V_a+ V_a^*)\outer{\psi}} \leq \tr{M_aV_a^*V_a \outer{\psi}}^{1/2}. 
$$
Now consider a certificate $\Gamma$ of~\eqref{eq:h2_noncommuting} which has the monomials $\{M_a, M_aV_a\}_a$ in its indexing set. Then as $\Gamma \geq 0$, for each $a$ the submatrix
$$
\begin{blockarray}{ccc}
~ & M_a & M_aV_a \\
\begin{block}{c(cc)}
M_a 	& \tr{M_a \outer{\psi}} & \tr{M_a V_a \outer{\psi}} \\
M_a V_a & \tr{M_a V_a^* \outer{\psi}} & \tr{M_a V_a^*V_a \outer{\psi}} \\
\end{block}
\end{blockarray}
$$
is positive semidefinite. Summing over $a$, the fact that each submatrix is PSD implies
\begin{equation*}
\begin{pmatrix}
\sum_a \tr{M_a \outer{\psi}} & \sum_a\tr{M_a V_a \outer{\psi}} \\
\sum_a\tr{M_a V_a^* \outer{\psi}} & \sum_a\tr{M_a V_a^*V_a \outer{\psi}}
\end{pmatrix} \geq 0.
\end{equation*}
By Lemma~\ref{lem:schur_complement} and the fact that $\sum_a \tr{M_a \outer{\psi}} = 1$ this implies that
\begin{equation}\label{eq:cauchy_from_psd}
\begin{aligned}
\sum_a\tr{M_a V_a^*V_a \outer{\psi}} &\geq (\sum_a\tr{M_a V_a^* \outer{\psi}})(\sum_a \tr{M_a V_a \outer{\psi}}) \\
&= (\sum_a\tr{M_a V_a \outer{\psi}})^2
\end{aligned}
\end{equation}
which is exactly the Cauchy-Schwarz relation. The final line follows from the fact that if $\Gamma$ is a real\footnote{We can always assume that $\Gamma$ is real and symmetric as if $\Gamma$ is a certificate then so is $(\Gamma + \overline{\Gamma})/2$, where $\overline{\Gamma}$ denotes the entrywise complex conjugate of $\Gamma$.} symmetric matrix then $\tr{M_a V_a \outer{\psi}} = \tr{M_a V_a^* \outer{\psi}}$. Thus, optimizing over such certificates we will always have
$$
\sum_a \tr{M_a\frac{V_a + V_a^*}{2} \outer{\psi}} \leq \left(\sum_a \tr{M_a V_a^* V_a \outer{\psi}}\right)^{1/2}.
$$
Now suppose $\Gamma_1$ is a certificate for~\eqref{eq:hmin_noncommuting} and $\Gamma_2$ is a certificate for~\eqref{eq:h2_noncommuting} which implies the Cauchy-Schwarz relation above. Then if for each monomial of the form $X W_a$ in the indexing set of $\Gamma_1$ we add a corresponding monomial $X V_a^* V_a$ to the indexing set of $\Gamma_2$ we will always have  
\begin{align*}
\max_{\Gamma_2} \sum_a \tr{M_a\frac{V_a + V_a^*}{2} \outer{\psi}} \leq&
\max_{\Gamma_2} \left(\sum_a \tr{M_a V_a^* V_a \outer{\psi}} \right)^{1/2} \\
\leq& \max_{\Gamma_1} \left(\sum_a \tr{M_aW_a \outer{\psi}}\right)^{1/2} .
\end{align*}
For example, when computing the plots from the main text we relaxed the $H_{\min}$ computations to the second level of the hierarchy. Then a sufficient relaxation for the $\imHup{2}$ computations is the second level of the hierarchy together with monomials $\{M_{a|x} V_c^* V_c\}_{a,x,c} \cup \{N_{b|y} V_c^* V_c\}_{b,y,c}$ where $\{M_{a|x}\}_{a,x}$ are operators representing Alice's measurements and $\{N_{b|y}\}_{b,y}$ are operators representing Bob's measurements.

Let us now consider the case of $\imHup{4/3}(A|E)$ from which the general case of $\imHup{\alpha_k}(A|E)$ follows readily. For this optimization we have additional operator inequalities
\begin{equation}
V_{1,a}^*V_{1,a} \leq \frac{V_{2,a} + V^*_{2,a}}{2}
\end{equation}
for each $a \in \cA$. Operator inequalities are imposed within the NPA hierarchy via localizing matrices (cf.~\eqref{eq:localizing_matrix}). That is, we take a collection of monomials $\cW_{\mathrm{loc}} = \{X_1, \dots X_k\}$ indexing a localizing matrix $\Gamma^{\mathrm{loc}} \geq 0$ whose $(X_i, X_j)$ entry corresponds to 
\begin{equation}
\tr{X_i^* \left(\frac{V_{2,a} + V^*_{2,a}}{2} - V_{1,a}^*V_{1,a}\right) X_j \outer{\psi}},
\end{equation}
for each $X_i,X_j \in \cW$. If the monomials $\{M_a\}$ corresponding to Alice's measurement operators are included in this localizing set $\cW_{\mathrm{loc}}$ then $\Gamma^{\mathrm{loc}} \geq 0$ enforces that
\begin{equation}\label{eq:loc_matrix_ineq}
\Gamma^{\mathrm{loc}}_{(M_a, M_a)} = \tr{M_a \left(\frac{V_{2,a} + V^*_{2,a}}{2} - V_{1,a}^*V_{1,a}\right) \outer{\psi}} \geq 0.
\end{equation}
By linearity of the trace this implies that $\tr{M_a \frac{V_{2,a} + V^*_{2,a}}{2} \outer{\psi}} \geq \tr{M_a V_{1,a}^*V_{1,a} \outer{\psi}}$. As in the above example for $\imHup{2}(A|E)$, if we add enough monomials to the indexing set of the certificate $\Gamma$ we can enforce Cauchy-Schwarz relations (cf.~\eqref{eq:cauchy_from_psd}). The Cauchy-Schwarz relation allows us to conclude that
\begin{equation}
\max_{\Gamma} \sum_a \tr{M_a\frac{V_{1,a} + V_{1,a}^*}{2} \outer{\psi}} \leq \max_{\Gamma} \left(\sum_a \tr{M_a V_{1,a}^* V_{1,a} \outer{\psi}}\right)^{1/2}
\end{equation}
and if we have sufficient monomials indexing the localizing matrices we can further conclude that
\begin{equation}
\max_{\Gamma} \left(\sum_a \tr{M_a V_{1,a}^* V_{1,a} \outer{\psi}}\right)^{1/2} \leq \max_{\Gamma} \left(\sum_a \tr{M_a \frac{V_{2,a} + V_{2,a}^*}{2} \outer{\psi}}\right)^{1/2}
\end{equation}
which is the objective function for $\imHup{2}(A|E)$.\footnote{We can move the $\max$ inside the exponentiation as $t \mapsto t^{1/2}$ is monotonic. Furthermore the exponent can be taken outside of the logarithm to cancel with the extra multiplicative factor of $2$ that $\imHup{4/3}$ has.} For general $\imHup{\alpha_k}$ this procedure can be repeated, including enough monomials in the certificate to enforce all of the Cauchy-Schwarz relations and for each operator inequality adding enough monomials to its corresponding localizing matrix to enforce the tracial inequalities of the form~\eqref{eq:loc_matrix_ineq}. 

\begin{remark}
	It is important that all necessary monomials are included. For example, it is common when certain variables in the optimization form a $n$-outcome POVM to remove one of them from the indexing set, e.g., defining the final element as $\id - M_1 - M_2 - \dots - M_{n-1}$. However, if this is done for the $\{M_a\}$ that appear in the objective function of $\imHup{\alpha_k}(A|E)$ then this will result in suboptimal rates as the relevant Cauchy-Schwarz relations will not be imposed. 
\end{remark}
\section{From SDPs to min-tradeoff functions}\label{app:sdp2tradeoff}
As noted in the main text, solutions to our device-independent optimizations may be combined with the entropy accumulation theorem~\cite{DFR,DF} in order to prove security of the respective device-independent protocols~\cite{ADFRV,ARV}. The entropy accumulation theorem specifies that, under reasonable assumptions, the total smooth min-entropy of a large system can be lower bounded by the total von Neumann entropy of its subsystems minus some correction term that scales sublinearly in the number of subsystems, i.e. 
\begin{equation}
H_{\min}^{\epsilon}(A_1^n B_1^n | X_1^n Y_1^n E) > \sum_{i=1}^n H(A_iB_i|X_iY_iE) - O(\sqrt{n}),
\end{equation}
where $F_1^n = F_1F_2\dots F_n$. The total smooth min-entropy characterizes the number of random uniform bits that can be extracted from $A_1^nB_1^n$ and so operationally corresponds to the length of the raw secret key for QKD (before losses due to error correction are taken into account) or the amount of gross uniform randomness acquired in randomness expansion. In order to use the entropy accumulation theorem to prove security of the protocol, one is required to construct \emph{min-tradeoff functions}. Recall that these are functions which lower bound the quantities $H(A_iB_i|X_iY_iE)$ in terms of some expected values of some statistical test $C:\cA\cB\cX\cY \rightarrow \cC$, e.g. an expected Bell-inequality violation. In the following we will show how it is possible to extract min-tradeoff functions directly from the solutions to our device-independent optimizations. 

Suppose we have a primal SDP of the following form 
\begin{equation}\label{eq:sdp_primal}
\begin{aligned}
p^*(b) := &\sup_{X}\quad \tr{C\,X} \\
& \quad\mathrm{s.t.} \quad \tr{F_i X} \geq b_i \qquad \text{for all } i = 1,\dots r \\
& \quad \qquad \,\, X \geq 0
\end{aligned}
\end{equation}
where $C, F_1, \dots, F_r$ are real symmetric matrices and $b_i \in \RR$. In the context with which we are concerned $X$ would correspond to a moment matrix of the NPA hierarchy and the inequality constraints impose the various constraints of the relaxation as well as the statistical constraints, e.g. a Bell-inequality violation. Note that we can impose equality constraints via two inequality constraints, i.e. $a \geq b$ and $-a \geq - b$ together imply $ a = b$. We chose to use the primal form with inequality constraints as this is how we implemented the SDPs, a similar computation could be done for an SDP with equality constraints. 

The dual of this optimization problem can be expressed as 
\begin{equation}\label{eq:sdp_dual}
\begin{aligned}
d^*(b) := &\inf_{\lambda_i \leq 0}\quad \sum_i \lambda_i b_i \\
& \quad\mathrm{s.t.} \quad C - \sum_i \lambda_i F_i - Y \leq 0 \\
& \quad \qquad \,\, Y \leq 0.
\end{aligned}
\end{equation}
Both the primal and the dual programs are parameterized by the constraint vector $b = (b_1,\dots,b_r)$. We will now show that we can use any feasible point of the dual program parameterized by $b$ to bound the optimal solution to the primal program parameterized by some other constraint vector $\hat{b} \in \RR^r$. Let $(\lambda, Y)$ be a feasible point of \eqref{eq:sdp_dual} when parameterized by the constraint vector $b$ and let $\widehat{X}$ be a feasible point of~\eqref{eq:sdp_primal} when parameterized by the constraint vector $\hat{b}$. Then we have
\begin{equation}
\begin{aligned}
0 &\geq \tr{(C - \sum_i \lambda_i F_i -Y)\widehat{X}} \\
&\geq \tr{C\widehat{X}}  - \sum_i \lambda_i\tr{F_i \widehat{X}} \\
&\geq \tr{C\widehat{X}}  - \sum_i \lambda_i \hat{b}_i.
\end{aligned}
\end{equation}
Thus, taking the supremum over all feasible $\widehat{X}$ we have $\sum_i \lambda_i \hat{b}_i \geq p^*(\hat{b})$. 

In the context of our device-independent optimizations we only need to vary certain parts of the constraint vector, i.e. the parts that correspond to the values of the statistical test. Therefore we can order the constraint vector such that it partitions into two smaller constraint vectors $b_{\mathrm{fix}}$ and $b_{\mathrm{var}}$ which are the fixed and varying parts of the full constraint vector respectively. We can also then partition the dual solution vector $\lambda = (\lambda_{\mathrm{fix}}, \lambda_{\mathrm{var}})$ in the same way. Writing $\alpha = \lambda_{\mathrm{fix}}  \cdot b_{\mathrm{fix}}$ we have that the dual solution provides us with an affine function $g(\hat{b}) := \alpha + \lambda_{\mathrm{var}} \cdot \hat{b}_{\mathrm{var}}$ which is always an upper bound on the primal program, $g(\hat{b}) \geq p^*(\hat{b})$. 

Let us return to the task of constructing min-tradeoff functions. Recall that a statistical test is some function $C: \cA\cB\cX\cY \rightarrow \cC$. Given a distribution $q:\cC \rightarrow [0,1]$, we say a strategy $(Q_A, Q_B, E, \ket{\psi}, \{M_{a|x}\}, \{N_{b|y}\})$ is compatible with the statistics $q$ if for all $c \in \cC$ we have 
\begin{equation}
\sum_{abxy: C(a,b,x,y) = c} \mu(x,y)p(a,b|x,y) = q(c),
\end{equation}
where $\mu$ is some probability distribution on $\cX\cY$. 
Then a function $f: \cP(\cC) \rightarrow \RR$ is a global min-tradeoff function for the statistical test $C$ if it satisfies 
\begin{equation}
f(q) \leq \inf_{\Sigma_C(q)} H(AB|XYE)
\end{equation}
where the infimum is taken over all post-measurement states of all finite-dimensional strategies that are compatible with statistics $q$. Similarly, we call $f$ a local min-tradeoff function if $f(q) \leq \inf_{\Sigma_C(q)} H(A|XE)$. 

Let $p^*_{\mathrm{NPA}}(q)$ be the optimal solution to an NPA relaxation of $\imQ{\alpha_k}^{\mathrm{DI}}$ (see \eqref{eq:di_h2_optimization} and \eqref{eq:h2_noncommuting}) with additional constraints of the form
$$
\pm \sum_{abxy: C(a,b,x,y) = c} \mu(x,y) \tr{(M_{a|x} \otimes N_{b|y} \otimes \id_E) \outer{\psi}}  \geq \pm q(c). 
$$
Then for any $k \in \NN$ and some fixed $(x_0, y_0) \in \cX\cY$ we have
\begin{equation}
\begin{aligned}
\inf_{\strat_C(q)} H(AB|XYE) &\geq \inf_{\strat_C(q)} \mu(x_0,y_0) H(AB|X=x_0,Y=y_0,E) \\
&= \mu(x_0,y_0)\inf_{\strat_C(q)} \frac{\alpha_{k}}{1-\alpha_{k}} \log \imQdi{\alpha_{k}} \\
&\geq \mu(x_0,y_0)\frac{\alpha_{k}}{1-\alpha_{k}} \log p^*_{\mathrm{NPA}}(q) \\
&\geq \mu(x_0,y_0)\frac{\alpha_{k}}{1-\alpha_{k}} \log(\alpha + \lambda \cdot q).
\end{aligned}
\end{equation}
That is, we can lower bound the von Neumann entropy with an iterated means entropy, solve the relaxed optimization of the iterated means entropy and then extract from the dual solution a lower bounding functional. Many device-independent protocols use a spot-checking procedure wherein the statistical test is performed infrequently and with high probability some fixed inputs $(x_0, y_0)$ are input to the devices. Hence the probability $\mu(x_0, y_0)$ will be close to one. In applications of the EAT the min-tradeoff functions are restricted to be affine functions of the statistics. However, as $-\log(\alpha + \lambda \cdot q)$ is a convex function of $q$ and so one can derive an affine lower bound by taking a first order Taylor expansion of $\mu(x_0,y_0)\frac{\alpha_{k}}{1-\alpha_{k}} \log(\alpha + \lambda \cdot q)$. The resulting function can then be used directly with the EAT. For example, we could repeat the analysis of~\cite{BRC}, which gave security proofs for randomness expansion using min-tradeoff functions derived from the min-entropy program, using our iterated means entropies. Given the comparisons between the iterated mean entropies and the min-entropy presented in the main text, redoing the security proofs in~\cite{BRC} with the iterated mean entropies would likely give substantial improvements on the finite round rates. For DI-QKD protocols one could look to adapt the analysis of \cite{ARV}, replacing the tradeoff function derived for the CHSH game~\cite{PABGMS} with tradeoff functions derived from our SDPs. 

\section{Additional plots}
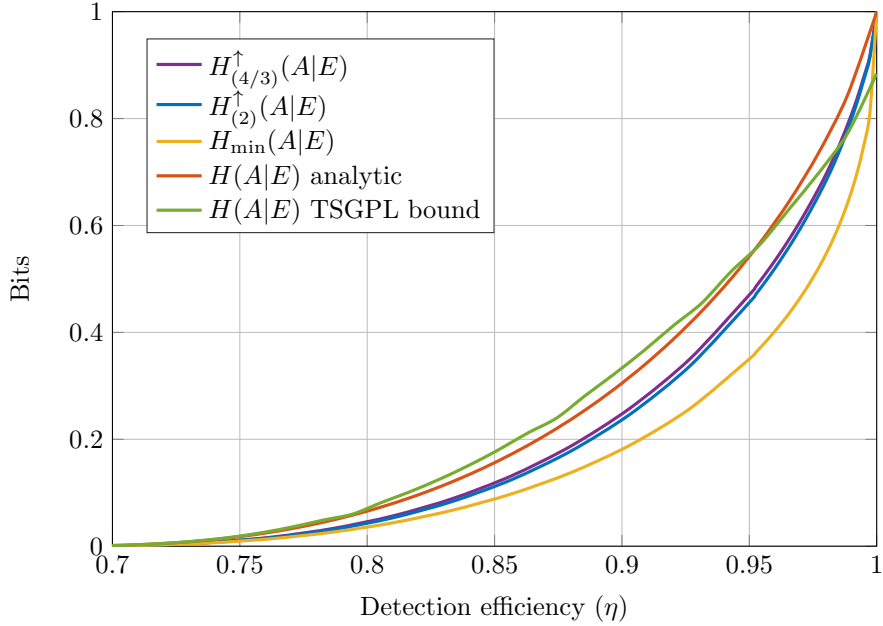
\begin{figure}
	\centering
	\definecolor{mycolor2}{rgb}{0.00000,0.44700,0.74100}%
	\definecolor{mycolor4}{rgb}{0.85000,0.32500,0.09800}%
	\definecolor{mycolor3}{rgb}{0.92900,0.69400,0.12500}%
	\definecolor{mycolor1}{rgb}{0.49400,0.18400,0.55600}%
	\definecolor{mycolor5}{rgb}{0.4660, 0.6740, 0.1880}%
	\begin{tikzpicture}
	
	\begin{axis}[%
	width=4in,
	height=2.8in,
	scale only axis,
	xmin=0.7,
	xmax=1.0,
	ymin=0,
	ymax=1.0,
	grid=major,
	xlabel={Detection efficiency ($\eta$)},
	ylabel={Bits},
	xtick={0.7, 0.75, 0.8, 0.85, 0.9, 0.95, 1.0},
	axis background/.style={fill=white},
	legend style={at={(0.5,0.95)},legend cell align=left, align=left, draw=white!15!black}
	]
	\addplot[smooth, color=mycolor1, line width=1.2pt] table[col sep=comma] {h43_local_DE_l2_tikz.dat};
	\addlegendentry{$\imHup{4/3}(A|E)$}
	\addplot[smooth, color=mycolor2, line width=1.2pt] table[col sep=comma] {h2_local_DE_l2_tikz.dat};
	\addlegendentry{$\imHup{2}(A|E)$}
	\addplot[smooth, color=mycolor3, line width=1.2pt] table[col sep=comma] {hmin_local_DE_l2_tikz.dat};
	\addlegendentry{$H_{\min}(A|E)$}
	\addplot[smooth, color=mycolor4, line width=1.2pt] table[col sep=comma] {chsh_analytic_de_tikz.dat};
	\addlegendentry{$H(A|E)$ analytic}
	\addplot[smooth, color=mycolor5, line width=1.2pt] table[col sep=comma] {ernest_local_de_tikz.dat};
	\addlegendentry{$H(A|E)$ TSGPL bound}
	\end{axis}
	\end{tikzpicture}%
	\caption{Comparison of lower bounds on $H(A|E)$ for quantum devices with inefficient detectors.}
	\label{fig:de_comparison_local}
\end{figure}

Results for the bounds on local randomness for 2-input 2-output devices constrained by their full conditional distribution are presented in Figure~\ref{fig:de_comparison_local}. As explained in the main text, for each detection efficiency we allow ourselves to optimize over some class of two-qubit systems to find a conditional distribution that maximizes the rate. We see a large difference between $\imHup{2}(A|E)$ and $H_{\min}(A|E)$. However, like in the corresponding plot for global randomness (cf. Figure~\ref{fig:de_comparison}), we see a negligible improvement on the randomness certified when comparing $\imHup{4/3}(A|E)$ and $\imHup{2}(A|E)$. Comparing with the analytical bound from~\cite{PABGMS} and the TSGPL bound from~\cite{TSGPL19}, we found our bounds are almost everywhere lower. An exception to this is in the regime of high detection efficiencies where our lower bounds converge to the optimum value of one and so surpass the TSGPL bound. 

\section{Proof of Proposition~\ref{prop:properties}}
For ease of reading recall that the iterated mean divergences are defined, for $k \in \NN$ and $\alpha_k = 1 + \frac{1}{2^k -1}$ as
\begin{equation}
\imD{\alpha_k}(\rho \| \sigma) := \frac{1}{\alpha_k-1} \log \imQ{\alpha_k}(\rho \| \sigma) \
\end{equation}
where
\begin{equation}
\begin{aligned}
\label{eq:def_imQ_repeat}
\imQ{\alpha_k}(\rho \| \sigma) := &\max_{V_1, \dots, V_k, Z} \,\,\alpha_k \tr{\rho \frac{(V_1 + V_1^{*})}{2}} - (\alpha_k - 1) \tr{\sigma Z} \\
& \quad\mathrm{s.t.} \quad\,\, V_1 + V_1^* \geq 0 \\
& \quad \quad \,\,\begin{pmatrix} \id & V_1 \\ V_1^* & \frac{(V_2 + V_2^*)}{2} \end{pmatrix} \geq 0 \quad \begin{pmatrix} \id & V_2 \\ V_2^* & \frac{(V_3 + V_3^*)}{2} \end{pmatrix} \geq 0 \quad \cdots \quad \begin{pmatrix} \id & V_k \\ V_k^{*} & Z \end{pmatrix} \geq 0 .
\end{aligned}
\end{equation}

Before we begin the proof of the proposition we make an observation that we can assume the support of all operators within the optimization is contained within the support of $\sigma$, i.e., $\sigma \gg Z$ and $\sigma \gg V_i$ for all $1 \leq i \leq k$. To see this consider the decomposition of the Hilbert space as $\cH = \supp(\sigma) \oplus \supp(\sigma)^\perp$. With respect to this decomposition we may write the operators in block matrix form as 
\begin{equation}
\rho = \begin{pmatrix}
\rho(0,0) & 0 \\
0 & 0
\end{pmatrix}, \quad 
\sigma = \begin{pmatrix}
\sigma(0,0) & 0 \\
0 & 0
\end{pmatrix}, \quad 
V_i = \begin{pmatrix}
V_i(0,0) & V_i(0,1) \\
V_i(1,0) & V_i(1,1)
\end{pmatrix}, \quad 
Z = \begin{pmatrix}
Z(0,0) & Z(0,1) \\
Z^*(0,1) & Z(1,1)
\end{pmatrix}.
\end{equation}
With this form the objective function may be written as
\begin{equation}
\alpha_k \tr{\rho(0,0) \frac{V_1(0,0) + V_1^*(0,0)}{2}} - (1-\alpha_k) \tr{\sigma(0,0) Z(0,0)}
\end{equation}
and so only depends on the restriction of the operators to the subspace $\supp(\sigma)$. Now the positive-semidefinite constraints in~\eqref{eq:def_imQ_repeat} may be rewritten as $V_i^* V_i \leq \frac{V_{i+1} + V_{i+1}^*}{2}$ for $1 \leq i \leq {k-1}$ and $V_k^* V_k \leq Z$. By direct computation we find that 
\begin{equation}
\frac{V_{i+1} + V_{i+1}^*}{2} - V_i^* V_i = 
\begin{pmatrix}
\frac{V_{i+1}(0,0) + V_{i+1}^*(0,0)}{2} - V_i^*(0,0)V_i(0,0) - V_i^*(1,0)V_i^*(1,0) & \quad*~\quad \\
* & \quad*~\quad
\end{pmatrix}
\end{equation}
and so $\frac{V_{i+1} + V_{i+1}^*}{2} - V_i^* V_i \geq 0 \implies \frac{V_{i+1}(0,0) + V_{i+1}^*(0,0)}{2} - V_i^*(0,0)V_i(0,0) - V_i^*(1,0)V_i(1,0) \geq 0 \implies \frac{V_{i+1}(0,0) + V_{i+1}^*(0,0)}{2} - V_i^*(0,0)V_i(0,0) \geq 0$. The final implication holds because $V_i^*(1,0)V_i(1,0) \geq 0$. Similarly, for the positive semidefinite constraint involving $Z$ we find $Z \geq V_k^* V_k \implies Z(0,0) \geq V_k^*(0,0) V_k(0,0)$. Finally $V_1 + V_1^* \geq 0 \implies V_1(0,0) + V_1^*(0,0) \geq 0$. Thus, denoting the projector onto the subspace $\supp(\sigma)$ by $\Pi$, we have that for any feasible point $(V_1,  \dots, V_k, Z)$, the point $(\Pi V_1\Pi,  \dots,\Pi V_k\Pi,\Pi Z\Pi)$ is also feasible, obtains the same objective value and all operators have their support contained in $\supp(\sigma)$. We therefore assume henceforth that all operators in the optimization have their support contained within $\supp(\sigma)$. 
~\\
~\\
{\noindent{\textbf{Property~1. Rescaling}}~\\
For any $\beta > 0$ we have $\begin{pmatrix}
	A & B \\
	B^* & C
\end{pmatrix} \geq 0 \iff \begin{pmatrix}
A & \beta B \\
\beta B^* & \beta^2 C
\end{pmatrix} \geq 0$. It follows then that for any feasible point $(V_1, \dots,V_k,Z)$ of~\eqref{eq:def_imQ_repeat}, $(\beta V_1, \beta^2 V_2, \dots, \beta^{2^{k-1}}V_k, \beta^{2^k}Z)$ is another feasible point. This new feasible point has an objective value $\alpha_k \beta \tr{\rho \frac{(V_1 + V_1^{*})}{2}} - (\alpha_k - 1) \beta^{2^k} \tr{\sigma Z}$. Assuming that $\tr{\rho \frac{(V_1 + V_1^{*})}{2}} \geq 0$ and $\tr{\sigma Z}>0$,\footnote{As $V_1 + V_1^* \geq 0$ we have $\tr{\rho(V_1 + V_1^*)} \geq 0$. Furthermore, as $Z \geq 0$ and $Z \ll \sigma$ we have $\tr{\sigma Z} = 0 \iff Z=0$. However if $Z=0$ then it follows from the other constraints that we must also have $V_1 = V_2 = \dots = V_k = 0$ and in turn the objective value is trivially $0$. We also have that for any $c > 0$, the point $(c \id, 2 c^2 \id, \dots, 2^{2^{k-1}-1}c^{2^{k-1}} \id, 2^{2^k-1}c^{2^k} \id)$ is feasible with an objective value $\alpha_k c \tr{\rho} -(\alpha_k-1) 2^{2^{k}-1} c^{2^k} \tr{\sigma}$. Rearranging we find that we have a strictly positive objective value when we choose $ c < (2^{1-2^k}\frac{\alpha_k}{\alpha_k - 1} \frac{\tr{\rho}}{\tr{\sigma}})^{\frac{1}{2^k - 1}}$. Thus the choice of $Z=0$ is always suboptimal and we may also assume that $\tr{\sigma Z} > 0$.} we may maximize over the choice of $\beta > 0$ and we find a unique maximum occurring at 
\begin{equation}
\beta^* = \left( \frac{\alpha_k}{2^k (\alpha_k -1)} \frac{\tr{\rho \frac{(V_1 + V_1^{*})}{2}}}{\tr{\sigma Z}} \right)^{\frac{1}{2^k - 1}}.
\end{equation}
For this choice of $\beta$ the objective function simplifies to 
\begin{equation}
\frac{\tr{\rho \frac{(V_1 + V_1^{*})}{2}}^{\alpha_k}}{\tr{\sigma Z}^{\frac{1}{2^k-1}}}.
\end{equation}
Note that after this rewriting, rescaling the operators as before with some $\beta > 0$ does not change the objective value. Thus, we are free to rescale the operators so that $\tr{\sigma Z} = 1$. Therefore we can rewrite the optimization as
\begin{equation}
\begin{aligned}
\imQ{\alpha_k}(\rho \| \sigma) = &\max_{V_1, \dots, V_k, Z} \,\, \tr{\rho \frac{(V_1 + V_1^{*})}{2}}^{\alpha_k} \\
& \quad\mathrm{s.t.} \quad \,\, \tr{\sigma Z} = 1 \\
& \qquad \,\, V_1 + V_1^* \geq 0 \\
&\quad\,\, \begin{pmatrix} \id & V_1 \\ V_1^* & \frac{(V_2 + V_2^*)}{2} \end{pmatrix} \geq 0 \quad \begin{pmatrix} \id & V_2 \\ V_2^* & \frac{(V_3 + V_3^*)}{2} \end{pmatrix} \geq 0 \quad \cdots \quad \begin{pmatrix} \id & V_k \\ V_k^{*} & Z \end{pmatrix} \geq 0 .
\end{aligned}
\end{equation}
~\\
~\\
{\noindent{\textbf{Property~2a. Dual form (a)}}~\\
We start by establishing the following dual form, which is not included in the statement for brevity:
\begin{equation}\label{eq:dual1_main}
\begin{aligned}
\imQ{\alpha_k}(\rho \| \sigma) = &\min_{A_1, \dots, A_k, C_1, \dots, C_{k} } \sum_{i=1}^k \tr{A_i} \\
& \qquad \qquad\mathrm{s.t.} \quad C_1 \geq \rho \\
& \qquad \qquad \quad \,\, \begin{pmatrix} A_1 & \frac{\alpha_k}{2} C_1 \\ \frac{\alpha_k}{2} C_1 & C_2 \end{pmatrix} \geq 0 \quad \begin{pmatrix} A_2 & \frac{C_2}{2} \\ \frac{C_2}{2} & C_3 \end{pmatrix} \geq 0 \qquad \cdots \qquad \begin{pmatrix} A_k & \frac{C_{k}}{2} \\ \frac{C_{k}}{2} & \frac{1}{2^k-1} \sigma \end{pmatrix} \geq 0 \ .
\end{aligned}
\end{equation}
Introducing the dual variables $\begin{pmatrix}
A_i & B_i \\
B_i^* & C_{i+1} 
\end{pmatrix}$ for $1 \leq i \leq k$ for the positive-semidefinite constraints and the dual variable $C_1$ for the constraint $V_1 + V_1^* \geq 0$ we can write the Lagrangian of the problem~\eqref{eq:def_imQ_repeat} as 
\begin{equation}
\begin{aligned}
L &= \alpha_k \tr{\rho \frac{(V_1 + V_1^{*})}{2}} - (\alpha_k - 1) \tr{\sigma Z} + \tr{(V_1 + V_1^*)C_1} \\
&+ \tr{A_1 + B_1 V_1^* + B_1^* V_1 + C_2(V_2 + V_2^*)/2} + \dots + \tr{A_k + B_k V_1^* + B_k^* V_1 + C_{k+1} Z} \\
&= \sum_{i=1}^k \tr{A_i} + \tr{V_1 (\tfrac{\alpha_k}{2} \rho + C_1 + B_1^*) +  V_1^* (\tfrac{\alpha_k}{2} \rho + C_1 + B_1)} + \dots + \tr{V_k (\tfrac{1}{2} C_{k} + B_k^*) +  V_{k}^* (\tfrac{1}{2} C_{k} + B_k)} \\
&+ \tr{Z(\tfrac12 C_{k+1} - (\alpha_k - 1) \sigma)} \\
&= \sum_{i=1}^k \tr{A_i} + 2 \mathscr{R}\left(\tr{V_1 (\tfrac{\alpha_k}{2} \rho + C_1 + B_1^*)}\right) + \dots + 2 \mathscr{R}\left(\tr{V_k (\tfrac{1}{2} C_{k} + B_k^*)}\right) + \tr{Z(\tfrac12 C_{k+1} - (\alpha_k - 1) \sigma)}
\end{aligned}
\end{equation}
where for the third equality we used the identity $ \tr{X + X^*} = 2 \mathscr{R}(\tr{X})$. Now if we take a maximization over the variables $V_1, \dots, V_k$ and $Z$, we find that the Lagrangian is finite only if $C_1 + \tfrac{\alpha_k}{2} \rho  + B_1^* = 0$, $B_i = -\tfrac12 C_{i-1}$ for $2\leq i \leq k$ and $C_k = (\alpha_k - 1)\sigma$. Note that the condition $C_1 + \tfrac{\alpha_k}{2} \rho + B_1^* = 0$ can be rewritten as $-B^*_1 \geq \tfrac{\alpha_k}{2} \rho$ as $C_1$ does not appear elsewhere. We relabel $-B^*_1$ to $\tfrac{\alpha_k}{2} C_1$. Also, note that it follows from Lemma~\ref{lem:schur_complement} that $\begin{pmatrix}
A & -B \\
-B^* & C
\end{pmatrix} \geq 0 \iff \begin{pmatrix}
A & B \\
B^* & C 
\end{pmatrix} \geq 0$. Therefore we can write the dual problem as 
\begin{equation}\label{eq:dual1}
\begin{aligned}
\imQ{\alpha_k}(\rho \| \sigma) = &\min_{A_1, \dots, A_k, C_1, \dots, C_{k} } \sum_{i=1}^k \tr{A_i} \\
& \qquad \qquad\mathrm{s.t.} \quad C_1 \geq \rho \\
& \qquad \quad \,\, \begin{pmatrix} A_1 & \frac{\alpha_k}{2} C_1 \\ \frac{\alpha_k}{2} C_1 & C_2 \end{pmatrix} \geq 0 \quad \begin{pmatrix} A_2 & \frac{C_2}{2} \\ \frac{C_2}{2} & C_3 \end{pmatrix} \geq 0 \qquad \cdots \qquad \begin{pmatrix} A_k & \frac{C_{k}}{2} \\ \frac{C_{k}}{2} & (\alpha_k -1) \sigma \end{pmatrix} \geq 0 \ .
\end{aligned}
\end{equation}

It remains to show that we have strong duality. In order to show this we observe that for any $c>0$ the assignment $V_1 = c \id$, $V_2 = 2 c^2 \id$, $\dots, V_k =2^{2^{k-1}-1} c^{2^{k-1}} \id$ and $Z = 2^{2^k - 1} c^{2^k} \id $ constitutes a strictly feasible point of the primal program. In the dual problem, for $2\leq i \leq k-1$ the constraints $\begin{pmatrix}
A_i & \tfrac12 C_{i} \\
 \tfrac12 C_{i} & C_{i+1}
\end{pmatrix} \geq 0$ have a strictly feasible assignment $C_i = C_{i+1} = 2 \id$ and $A_i = \tfrac{c}{2} \id$ for any $c > 1$. Then the assignment $A_1 = c \tfrac{\alpha_k^2}{2}$ and $C_1 = 2 \id$ satisfies the first positive semidefinite constraint and $C_1 \geq \rho$ strictly. Now recall that we may assume that we work in the subspace $\supp(\sigma)$ and so we have $\sigma > 0$. Therefore, the assignment $A_k = \frac{c}{(\alpha_{k}-1)} \sigma^{-1}$ satisfies the final constraint strictly. As we have demonstrated strictly feasible points to both the primal and the dual problems, it follows that we have strong duality.
~\\
~\\
{\noindent{\textbf{Property~2b. Dual form (b)}}~\\
Firstly, note that it follows from Lemma~\ref{lem:schur_complement} that for any $\beta > 0$ we have $\begin{pmatrix}
A & \beta B \\
\beta B^* & C  
\end{pmatrix} \geq 0 \iff 
\begin{pmatrix}
\tfrac{1}{\beta}A &  B \\
B^* & \tfrac{1}{\beta}C  
\end{pmatrix} \geq 0$. Then we can rewrite the block matrix constraints of the dual problem~\eqref{eq:dual1} as
\begin{equation*}
\begin{pmatrix}
\tfrac{2}{\alpha_{k}} A_1 & C_1 \\
C_1 & \tfrac{2}{\alpha_k} C_2 
\end{pmatrix} \geq 0 \qquad
\begin{pmatrix}
\tfrac{4}{\alpha_{k}} A_2 & \tfrac{4}{\alpha_{k}} \tfrac12 C_2 \\
\tfrac{4}{\alpha_{k}} \tfrac12 C_2 & \tfrac{4}{\alpha_k} C_3 
\end{pmatrix} \geq 0 \qquad \dots \qquad
\begin{pmatrix}
\tfrac{2^k}{\alpha_{k}} A_k & \tfrac{2^k}{\alpha_{k}} \tfrac12 C_{k} \\
\tfrac{2^k}{\alpha_{k}} \tfrac12 C_{k} & \tfrac{2^k}{\alpha_k} (\alpha_k-1) \sigma 
\end{pmatrix} \geq 0.
\end{equation*}
Making the change of variables $\widehat{A}_i = \tfrac{2^i}{\alpha_k} A_i$ and $\widehat{C}_i = \tfrac{2^i}{\alpha_k} C_i$ for $2 \leq i \leq k$, we find that the dual program~\eqref{eq:dual1} is equivalent to 
\begin{equation}\label{eq:dual2}
\begin{aligned}
&\min_{A_1, \dots, A_k, C_1, \dots, C_{k} } \frac{1}{2^k - 1}\sum_{i=1}^k 2^{k-i}\tr{A_i} \\
& \qquad \qquad\mathrm{s.t.} \quad C_1 \geq \rho \\
& \qquad \qquad \quad \begin{pmatrix} A_1 & C_1 \\ C_1 & C_2 \end{pmatrix} \geq 0 \quad \begin{pmatrix} A_2 & C_2 \\ C_2 & C_3 \end{pmatrix} \geq 0 \qquad \cdots \qquad \begin{pmatrix} A_k & C_{k} \\ C_{k} & \sigma \end{pmatrix} \geq 0 \ ,
\end{aligned}
\end{equation}
where we also used the fact that the coefficient of $\sigma$ simplifies as $\tfrac{2^k}{\alpha_k} (\alpha_k-1) = 1$.
~\\
~\\
{\noindent{\textbf{Property~2c. Dual form (c)}}~\\
We now derive the third dual form from the second dual form~\eqref{eq:dual2}. Firstly, let $\gamma_1 > 0$ and note that it follows from Lemma~\ref{lem:schur_complement} that for any feasible point $(A_1,\dots,A_k,C_1,\dots,C_{k})$ of~\eqref{eq:dual2}, 
$$
(\gamma_1 A_1, \tfrac{1}{\gamma_1^2}A_2,A_3,\dots,A_k, C_1, \tfrac{1}{\gamma_1} C_2, \dots,C_{k})
$$ 
is also a feasible point. By setting $\gamma_1 = \left(\frac{\tr{A_2}}{\tr{A_1}}\right)^{1/3}$ we have $\tr{\gamma_1 A_1} = \tr{\tfrac{1}{\gamma_1} A_2}$. Furthermore, we have for this choice of $\gamma_1$ that 
\begin{equation*}
\begin{aligned}
2 \tr{\gamma_1 A_1} + \tr{\tfrac{1}{\gamma_1}A_2} &= 3 \tr{A_1}^{2/3} \tr{A_2}^{1/3} \\
&\leq 2 \tr{A_1} + \tr{A_2}, 
\end{aligned}
\end{equation*}
where the second line follows from the arithmetic-geometric mean inequality. This shows that for any feasible point we can transform it to another feasible point such that $\tr{A_1} = \tr{A_2}$ and the objective value does not increase under the transformation.

We shall now demonstrate that we can inductively transform any feasible point into another such that the objective value does not increase and the transformed point satisfies $\tr{A_1} = \tr{A_2} = \dots = \tr{A_k}$. Suppose we have a feasible point $(A_1, \dots A_k, C_1, \dots, C_{k})$ such that $\tr{A_1} = \tr{A_2} = \dots = \tr{A_{i-1}}$ for some $2 \leq i \leq k $. Then by Lemma~\ref{lem:schur_complement} the point 
\begin{equation*}
(\gamma_i A_1, \gamma_i A_2, \dots, \gamma_i A_{i-1}, \gamma_i^{-2(2^i -1)}A_i, A_{i+1}, \dots, A_k, C_1, \gamma_i^{-1} C_2, \gamma_i^{-3} C_3, \dots, \gamma_i^{-(2^i-1)}C_{i}, C_{i+1}, \dots, C_{k})
\end{equation*}
is also feasible. By setting $\gamma_i = \left(\frac{\tr{A_i}}{\tr{A_1}}\right)^{\frac{1}{2^{i+1} - 1}}$ we get $\tr{\gamma_i A_1} = \tr{\gamma_i^{-2(2^i -1)}A_i}$. Furthermore, for this choice of $\gamma_i$ we have 
\begin{equation*}
\begin{aligned}
2^i \tr{\gamma_i A_1} + 2^{i-1}\tr{\gamma_i A_2} + \dots + 2 \tr{\gamma_i A_{i-1}} + \tr{\gamma_i^{-2(2^i -1)}A_i} &= 2 (2^i -1) \tr{\gamma_i A_1} + \tr{\gamma_i^{-2(2^i -1)}A_i} \\ 
&= (2^{i+1} -1) \tr{A_1}^{1- \frac{1}{2^{i+1}-1}}\tr{A_i}^{\frac{1}{2^{i+1}-1}} \\ 
&\leq 2 (2^i - 1) \tr{A_1} + \tr{A_i} \\
&= 2^i \tr{A_1} + 2^{i-1} \tr{A_2} + \dots + \tr{A_i},
\end{aligned}
\end{equation*}
where on the first line we used $\tr{A_1} = \tr{A_2} = \dots = \tr{A_{i-1}}$, the second line we substituted in our choice of $\gamma_i$ and the third line is another application of the arithmetic-geometric mean inequality. This shows that the objective value of the transformed point is no larger than that of the original point. It then follows by induction that we can transform any feasible point of~\eqref{eq:dual2} into one which satisfies $\tr{A_1}=\tr{A_2}= \dots = \tr{A_k}$ without increasing the objective value. Finally, noting that $\tfrac{1}{2^k - 1} \sum_i 2^{k-i} \tr{A_1} = \tr{A_1}$ we find that we can rewrite~\eqref{eq:dual2} as
\begin{equation}\label{eq:dual3}
\begin{aligned}
&\min_{A_1, \dots, A_k, C_1, \dots, C_{k} } \tr{A_1} \\
& \qquad \qquad\mathrm{s.t.} \quad \tr{A_1} = \tr{A_2} =\dots = \tr{A_k} \\
& \qquad \qquad \qquad C_1 \geq \rho \\
&\qquad \qquad \qquad \begin{pmatrix} A_1 & C_1 \\ C_1 & C_2 \end{pmatrix} \geq 0 \quad \begin{pmatrix} A_2 & C_2 \\ C_2 & C_3 \end{pmatrix} \geq 0 \qquad \cdots \qquad \begin{pmatrix} A_k & C_{k} \\ C_{k} & \sigma \end{pmatrix} \geq 0 \ ,
\end{aligned}
\end{equation}
~\\
~\\
~{\noindent{\textbf{Property~2d. Dual form (d)}}
	We derive the final dual form from the third dual form~\eqref{eq:dual3} -- an alternative dual form could be derived by starting at~\eqref{eq:dual2}. Consider any feasible point $(A_1,\dots A_k, C_1, \dots C_k)$ of \eqref{eq:dual3}. By Lemma~\ref{lem:psd_to_mgm} we know that 
	\begin{equation*}
	\begin{pmatrix}
	A_i & C_i \\
	C_i & C_{i+1} 
	\end{pmatrix} \geq 0 \implies C_i \leq A_i \# C_{i+1}.
	\end{equation*}
	Therefore the block matrix constraints of~\eqref{eq:dual3} imply the operator inequalities
	\begin{equation*}
	C_1 \leq A_1 \# C_2 \quad C_2 \leq A_2 \# C_3 \quad \dots \quad C_k \leq A_k \# \sigma.
	\end{equation*}
	Using the fact that if $C \leq D$ then $A \# C \leq A \# D$, we can combine these inequalities together with $\rho \leq C_1$ to conclude that any feasible point of~\eqref{eq:dual3} is also a feasible point of the optimization problem
	\begin{equation}\label{eq:dual4}
	\begin{aligned}
	&\min_{A_1, \dots, A_k} \tr{A_1} \\
	& \qquad \qquad\mathrm{s.t.} \quad  \tr{A_1} = \tr{A_2} = \dots = \tr{A_k} \\
	&\qquad \qquad \qquad \rho \leq  A_1 \# ( A_2 \# (\dots \# (A_k \# \sigma) \dots) ).
	\end{aligned}
	\end{equation}
	Moreover, the objective value remains unchanged. Now consider a feasible point $(A_1, \dots A_k)$ of~\eqref{eq:dual4}. As $\begin{pmatrix}
	A & A\#B \\
	A\#B & B
	\end{pmatrix} \geq 0$ it follows that by choosing $C_i = A_i \# A_{i+1} \dots \# A_k \# \sigma$ for each $i = 1,\dots, k$ that $(A_1, \dots, A_k, C_1, \dots, C_k)$ is a feasible point of~\eqref{eq:dual3} with the same objective value. Therefore~\eqref{eq:dual3} and~\eqref{eq:dual4} are equal. 
~\\
~\\
{\noindent{\textbf{Property~3. Submultiplicativity}}~\\
Let $(A_1, \dots, A_k,C_1,\dots,C_k)$ be the optimal point of the optimization~\eqref{eq:dual3} for the parameter pair $(\rho, \sigma)$ and let $(\widehat{A}_1, \dots, \widehat{A}_k,\widehat{C}_1,\dots,\widehat{C}_k)$ be the optimal point of~\eqref{eq:dual3} for the parameter pair $(\widehat{\rho}, \widehat{\sigma})$. Then $(A_1 \otimes \widehat{A}_1,\dots,A_k \otimes \widehat{A}_k,C_1 \otimes \widehat{C}_1, \dots, C_k \otimes \widehat{C}_k)$ is a feasible point of~\eqref{eq:dual3} for the pair $(\rho \otimes \widehat{\rho}, \sigma \otimes \widehat{\sigma})$. Moreover, we then have
\begin{equation*}
\begin{aligned}
\imQ{\alpha_k}(\rho \otimes \widehat{\rho} \| \sigma \otimes \widehat{\sigma}) 
&\leq \tr{A_1 \otimes \widehat{A}_1} \\
&=\tr{A_1}\tr{\widehat{A}_1} \\
&=\imQ{\alpha_k}(\rho \| \sigma) \imQ{\alpha_k}(\widehat{\rho} \| \widehat{\sigma}), 
\end{aligned}
\end{equation*}
and so $\imD{\alpha_k}(\rho \otimes \widehat{\rho} \| \sigma \otimes \widehat{\sigma}) \leq  \imD{\alpha_k}(\rho \| \sigma) + \imD{\alpha_k}(\widehat{\rho} \| \widehat{\sigma})$.
~\\
~\\
{\noindent{\textbf{Property~4. Relation to other R\'enyi divergences}}~\\
	Recall that $D^{\mathbb{M}}_{\alpha_k}(\rho \| \sigma) = \frac{1}{\alpha_k - 1} \log \max_{\omega > 0} \alpha_k \tr{\rho \omega} + (1-\alpha_k) \tr{\sigma \omega^{2^k}}$. Any $\omega > 0$ defines a feasible choice $V_i = \omega^{2^{i-1}}$ and $Z = \omega^{2^k}$. This gives us immediately $\imD{\alpha_k}(\rho \| \sigma) \geq D^{\mathbb{M}}_{\alpha_k}(\rho \| \sigma)$. Then by submultiplicativity, for any integer $n \geq 1$,
	\begin{align*}
	\imD{\alpha_k}(\rho \| \sigma) \geq \frac{1}{n} \imD{\alpha_k}(\rho^{\otimes n} \| \sigma^{\otimes n}) \\
	\geq \frac{1}{n} D^{\mathbb{M}}_{\alpha_k}(\rho^{\otimes n} \| \sigma^{\otimes n}) \ .
	\end{align*}
	Taking the limit as $n \to \infty$, we get the sandwiched R\'enyi divergence and so $\imD{\alpha_k}(\rho\|\sigma) \geq \widetilde{D}_{\alpha_{k}}(\rho \| \sigma)$~\cite{Tom15book}.
	~\\
	~\\
	{\noindent{\textbf{Property~5. Decreasing in $k$}~\\
	To show the fact that $\imD{\alpha_k}$ is decreasing in $k$, we write using the Cauchy-Schwarz inequality and the fact that $\tr{\rho} = 1$,
	\begin{align*}
	\imD{\alpha_k}(\rho \| \sigma) &= 2^k \log \max_{V_1, \dots, V_k, Z} \tr{\rho (V_1+V_1^*)/2} \\
	&\leq 2^k \log \max_{V_1, \dots, V_k, Z} \sqrt{\tr{\rho V_1^*V_1}} \\
	&\leq 2^k \log \max_{V_2, \dots, V_k, Z} \sqrt{\tr{\rho (V_2 + V_2^*)/2}} \\
	&= 2^{k-1} \log \max_{V_2, \dots, V_k, Z} \tr{\rho (V_2 + V_2^*)/2} \\
	&= \imD{\alpha_{k-1}}(\rho \| \sigma)
	\end{align*}
	where the third line follows from the operator inequality constraint $V_1^* V_1 \leq \frac{V_2+V_2^*}{2} $.
~\\
~\\
{\noindent{\textbf{Property~6. Data processing}}~\\
		Let $\cE^\dagger$ be the adjoint channel of some CPTP map $\cE: \Lin(A) \rightarrow \Lin(B)$. Note that $\cE^\dagger$ is unital and completely positive. Now consider the optimization
		\begin{align*}
		q=&\max_{W_1, \dots, W_k, Y} \left(\tr{\rho \frac{(\cE^\dagger(W_1) + \cE^\dagger(W_1)^{*})}{2}} \right)^{\alpha_k}  \\
		&\text{s.t.} \quad  \tr{\sigma \cE^\dagger(Y)} = 1  \\
		& \quad \cE^{\dagger}(W_1) + \cE^{\dagger}(W_1)^* \geq 0 \\
		&\quad \begin{pmatrix} \id & \cE^\dagger(W_1) \\ \cE^\dagger(W_1)^* & \frac{(\cE^\dagger(W_2) + \cE^\dagger(W_2)^*)}{2} \end{pmatrix} \geq 0 \quad \begin{pmatrix} \id & \cE^\dagger(W_2) \\ \cE^\dagger(W_2)^* & \frac{(\cE^\dagger(W_3) + \cE^\dagger(W_3)^*)}{2} \end{pmatrix} \geq 0 \quad \cdots \quad \begin{pmatrix} \id & \cE^\dagger(W_k) \\ \cE^\dagger(W_k)^{*} & \cE^\dagger(Y) \end{pmatrix} \geq 0 \ ,
		\end{align*}
		where the optimization is over linear operators on $B$. Identifying $V_i = \cE^{\dagger}(W_i)$ and $Z = \cE^\dagger(Y)$ we see that every feasible point for the above optimization defines a feasible point for the optimization $\imQ{\alpha_k}(\rho\|\sigma)$ with the same objective value. Therefore we must have $\imQ{\alpha_k}(\rho\|\sigma) \geq q$. Now as $\cE^{\dagger}$ is completely positive it also preserves adjoints, i.e., $\cE^{\dagger}(W^*) = \cE^{\dagger}(W)^*$. 
		Therefore, using the fact that $\cE^{\dagger}$ is also unital, we can rewrite $q$ as 
		\begin{align*}
		q=&\max_{W_1, \dots, W_k, Y} \left(\tr{\cE(\rho) \frac{(W_1 + W_1^{*})}{2}} \right)^{\alpha_k}  \\
		&\text{s.t.} \quad  \tr{\cE(\sigma) Y} = 1, \\
		&\quad \cE^{\dagger}(W_1 + W_1^*) \geq 0 \\
		&\quad (\cI_2 \otimes \cE^{\dagger})\begin{pmatrix} \id & W_1 \\ W_1^* & \frac{(W_2 + W_2^*)}{2} \end{pmatrix} \geq 0 \quad (\cI_2 \otimes \cE^{\dagger})\begin{pmatrix} \id & W_2 \\ W_2^* & \frac{(W_3 + W_3^*)}{2} \end{pmatrix} \geq 0 \quad \cdots \quad (\cI_2 \otimes \cE^{\dagger})\begin{pmatrix} \id & W_k \\ W_k^{*} & Y \end{pmatrix} \geq 0 \ .
		\end{align*}
		Writing
		\begin{align*}
		\imQ{\alpha_k}(\cE(\rho)\|\cE(\sigma)) =&\max_{W_1, \dots, W_k, Y} \left(\tr{\cE(\rho) \frac{(W_1 + W_1^{*})}{2}} \right)^{\alpha_k}  \\
		&\text{s.t.} \quad  \tr{\cE(\sigma) Y} = 1,  \\
		&\quad W_1 + W_1^* \geq 0 \\
		&\quad \begin{pmatrix} \id & W_1 \\ W_1^* & \frac{(W_2 + W_2^*)}{2} \end{pmatrix} \geq 0 \quad \begin{pmatrix} \id & W_2 \\ W_2^* & \frac{(W_3 + W_3^*)}{2} \end{pmatrix} \geq 0 \quad \cdots \quad \begin{pmatrix} \id & W_k \\ W_k^{*} & Y \end{pmatrix} \geq 0 \ ,
		\end{align*}
		we see that we must also have $q \geq \imQ{\alpha_k}(\cE(\rho)\|\cE(\sigma))$ as they have the same objective function but each feasible point of the latter is a feasible point of the former as $\cE^\dagger$ is completely positive. Hence, we have $\imQ{\alpha_k}(\rho\|\sigma) \geq q \geq \imQ{\alpha_k}(\cE(\rho)\|\cE(\sigma))$ and as $\frac{1}{\alpha_k - 1} \log(\cdot)$ is monotonically increasing for all $k \in \NN$ the result follows.
~\\
~\\
{\noindent{\textbf{Property~7. Reduction to classical divergence}}~\\
If $[\rho,\sigma] = 0$ then there exists a common eigenbasis of $\rho$ and $\sigma$, i.e. there exists an orthonormal basis $\{\ket{x}\}$ such that $\rho = \sum_x p_x \outer{x}$ and $\sigma = \sum_x q_x \outer{x}$ with $p_x, q_x \geq 0$ and $\sum_x p_x = \sum_x q_x = 1$. Let $\mathcal{P}: \Lin(\cH) \rightarrow \Lin(\cH)$ be the pinching map
\begin{equation*}
\mathcal{P}(A) = \sum_x \outer{x} A\outer{x}
\end{equation*}
defined by this common eigenbasis. Now consider any feasible point $(A_1,\dots A_k, C_1, \dots, C_{k})$ of the dual problem~\eqref{eq:dual2}. As the pinching map $\mathcal{P}$ is completely positive, $\rho = \mathcal{P}(\rho)$ and $\sigma = \mathcal{P}(\sigma)$, it follows that $(\mathcal{P}(A_1), \dots, \mathcal{P}(A_k), \mathcal{P}(C_1), \dots, \mathcal{P}(C_{k}))$ is another feasible point of the dual problem. Moreover, this new feasible point has the same objective value as the original point. Therefore, when $\rho$ and $\sigma$ commute we may assume that all variables in the optimization also commute. 

Now we know that~\cite[Proposition 3.3.4]{hiainotes}
$$
\begin{pmatrix}
A_1 & C_1 \\
C_1 & C_2
\end{pmatrix} \geq 0 \implies C_1 \leq A_1 \# C_2  = A_1^{1/2} C_2^{1/2}
$$
where the final equality holds as all operators are assumed to commute. Similarly, we have
$$
\begin{pmatrix}
A_2 & C_2 \\
C_2 & C_3
\end{pmatrix} \geq 0  \implies C_2 \leq A_2 \# C_3  = A_2^{1/2} C_3^{1/2}. 
$$ 
As all operators commute, these inequalities, together with $\rho \leq C_1$, imply that $\rho \leq A_1^{1/2} A_2^{1/4} C_2^{1/4}$. Repeating this for the remaining PSD constraints in the dual problem we find that $\rho \leq A_1^{1/2} \dots A_k^{1/2^k} \sigma^{1/2^k}$ or equivalently $\rho \sigma^{-1/2^k}\leq A_1^{1/2} \dots A_k^{1/2^k}$. Noting that $-\alpha_k/2^k = 1-\alpha_k$, by taking both sides of the inequality to the power of $\alpha_k$ we arrive at 
$$
\rho^{\alpha_k} \sigma^{1-\alpha_{k}} \leq A_1^{\alpha_k/2} \dots A_k^{\alpha_k/2^k}.
$$
It follows that
\begin{align*}
\tr{\rho^{\alpha_k} \sigma^{1-\alpha_{k}}} &\leq \tr{A_1^{\alpha_k/2} \dots A_k^{\alpha_k/2^k}} \\
&\leq \sum_{i=1}^{k} \frac{\alpha_{k}}{2^i} \tr{A_i} \\
&= \frac{1}{2^k - 1}\sum_{i=1}^k 2^{k-i}\tr{A_i},
\end{align*}
where the second line follows from the arithmetic-geometric mean inequality. Thus, when $[\rho, \sigma] = 0$ we know that $\imD{\alpha_k}(\rho \| \sigma) \geq \frac{1}{\alpha_k - 1} \log \tr{\rho^{\alpha_{k}} \sigma^{1- \alpha_{k}}}$. 

It remains to show that there always exists a feasible point that achieves this bound. For this we choose $A_1 = A_2 = \dots = A_k = \rho^{\alpha_k} \sigma^{1-\alpha_k}$. It can be verified that this choice satisfies the inequality
$$
\rho \leq A_1 \# ( A_2 \# (\dots \# (A_k \# \sigma) \dots) )
$$
as well as the other constraints of the dual form~\eqref{eq:dual4}. Therefore, there exists a feasible point of~\eqref{eq:dual4} achieving the lower bound $\tr{\rho^{\alpha_{k}} \sigma^{1- \alpha_k}}$ and so the result follows. 

\section{Additional Lemmas}

The following lemma provides a useful characterization of positive semidefiniteness for block matrices.
\begin{lemma}[Schur complement]\label{lem:schur_complement}
	Let $A, B, C \in \Lin(\cH)$. Then the following are all equivalent:
	\begin{enumerate}
		\item $\begin{pmatrix}
		A & B \\
		B^* & C
		\end{pmatrix} \geq 0$.
		\item $A \geq 0$, $(\id - A A^{-1})B = 0$ and $ C \geq B^* A^{-1} B$.
		\item $ C \geq 0$, $(\id - C C^{-1})B^* = 0$ and $A \geq B C^{-1}B^*$.
	\end{enumerate}
	Furthermore, if we restrict to positive-definite matrices then the following are equivalent:
	\begin{enumerate}
		\item $\begin{pmatrix}
		A & B \\
		B^* & C
		\end{pmatrix} > 0$.
		\item $A > 0$ and $ C > B^* A^{-1} B$.
		\item $ C > 0$ and $A > B C^{-1}B^*$.
	\end{enumerate}
\end{lemma}

The following lemma relates block positive semidefinite matrices to the matrix geometric mean. 
\begin{lemma}\label{lem:psd_to_mgm}
	Let $A, B \in \Pos(\cH)$ and $T \in \Her(\cH)$. Then $A\# B \geq T \iff \exists W \in \Her(\cH)$ such that $W \geq T$ and 
	\begin{equation}
	\begin{pmatrix}
	A & W \\
	W & B
	\end{pmatrix} \geq 0.
	\end{equation}
	\begin{proof}
		It is well-known that (see e.g., \cite[Proposition 3.3.4]{hiainotes}) that for $A, B \in \Pos(\cH)$ and $W \in \Her(\cH)$ then $\begin{pmatrix}
		A & W \\
		W & B
		\end{pmatrix} \geq 0 \implies A \# B \geq W$. Therefore if in addition $W \geq T$ we have $A \# B \geq T$. Additionally, they also show that $\begin{pmatrix}
		A & A\# B \\
		A \# B & B 
		\end{pmatrix} \geq 0$ and so the converse holds also.  
	\end{proof}
\end{lemma}
\end{document}